\documentclass[a4paper,10pt,notitlepage]{article}
\usepackage[utf8x]{inputenc}
\usepackage{amsfonts,amssymb,amsmath,amsthm,bbold,url,graphicx,psfrag}
\usepackage{slashed}

\usepackage{epstopdf}

\usepackage{tikz}
\usetikzlibrary{matrix,arrows,decorations.pathmorphing}

\numberwithin{equation}{section}

\DeclareMathOperator{\SO}{SO}

\DeclareMathOperator{\GL}{GL}
\DeclareMathOperator{\SU}{SU}
\DeclareMathOperator{\U}{U}

\DeclareMathOperator{\Sym}{Sym}

\DeclareMathOperator{\un}{\mathfrak{u}}
\DeclareMathOperator{\so}{\mathfrak{so}}
\DeclareMathOperator{\sgl}{sl}
\DeclareMathOperator{\gl}{\mathfrak{gl}}
\DeclareMathOperator{\su}{\mathfrak{su}}

\DeclareMathOperator{\res}{res}

\DeclareMathOperator{\tr}{tr}
\DeclareMathOperator{\sdet}{sdet}
\DeclareMathOperator{\col}{col}
\DeclareMathOperator{\row}{row}

\DeclareMathOperator{\wt}{wt}
\newcommand{\Yan}{Y}
\DeclareMathOperator{\ch}{ch}
\DeclareMathOperator{\sch}{sch}

\DeclareMathOperator{\Cas}{Cas}

\DeclareMathOperator{\Tab}{Tab}
\DeclareMathOperator{\STab}{STab}

\title{Supersymmetric holography on AdS$_3$}
\author{Constantin Candu and Matthias R.\ Gaberdiel \\
Institut f\"ur Theoretische Physik , ETH Z\"urich \\
CH-8093 Z\"urich, Switzerland}

\begin{document}
\begin{titlepage}
\maketitle
%
\begin{abstract}
The proposed duality between 
Vasiliev's supersymmetric higher spin theory on AdS$_3$ 
and the 't~Hooft limit of the 2d $\mathcal{N}=2$ superconformal 
Kazama-Suzuki models is analysed in detail.
In particular, we show that the partition functions of the two theories
agree in the large $N$ limit.
\end{abstract}

\tableofcontents

\end{titlepage}

\section{Introduction}

Theories containing an infinite number of (massless) higher spin 
currents are an interesting class of theories that lie in 
complexity somewhere between field and string theories. The first 
non-trivial examples where constructed about twenty-five years ago by 
Fradkin \& Vasiliev \cite{Fradkin:1986qy,Fradkin:1987ks}.
Recently, these theories have gained prominence
in the  context of the AdS/CFT correspondence since they are believed to be dual to 
free conformal theories \cite{Sundborg:2000wp,Witten,Mikhailov:2002bp,Sezgin:2002rt}. 
This offers the hope of finding simplified versions of the AdS/CFT duality. It may also open 
the way towards a proof of the AdS/CFT correspondence, at least in a specific regime;
for first attempts in this direction see 
\cite{Koch:2010cy,Douglas:2010rc,Jevicki:2011ss,Jevicki:2011aa}.

About ten years ago it was conjectured by Klebanov \& Polyakov 
\cite{Klebanov:2002ja} (see also \cite{Sezgin:2003pt} for a subsequent refinement) 
that a specific higher spin theory on AdS$_4$  \cite{Vasiliev:2003ev} (see for example
\cite{Vasiliev:1999ba,Bekaert:2005vh,Iazeolla:2008bp,Campoleoni:2009je} for reviews)
is dual to the large $N$ limit of the O($N$) vector model in $3$ dimensions; 
actually, there are four different versions of this duality, depending on whether
one considers the free or interacting O($N$) theory, and whether it is based on fermions
or bosons. During the last two years, highly non-trivial evidence in favour of this conjecture 
has been found. In particular, Giombi \& Yin managed to calculate some 3-point
functions of the higher spin theory on AdS$_4$, and showed that they
reproduce precisely those of the dual O($N$) vector model in the large $N$
limit \cite{Giombi:2009wh,Giombi:2010vg,Giombi:2011ya}. For the interacting
theory, the higher spin symmetry gets broken at finite $N$ \cite{Maldacena:2011jn},
but the symmetry may still play a useful role in determining the correlators of the theory.

The argument of \cite{Maldacena:2011jn} only applies to $3$d conformal field theories,
whereas in $2$ dimensions it is known that interacting higher spin theories 
(even with a finite number of degrees of freedom) exist, for example, the 
$W_N$ minimal models. A little while ago, it was shown that the asymptotic 
symmetry algebra of higher spin theories on AdS$_3$ 
\cite{Prokushkin:1998bq,Prokushkin:1998vn} lead to classical
$W_N$ or $W_\infty$ symmetry algebras 
\cite{Henneaux:2010xg,Campoleoni:2010zq,Gaberdiel:2011wb,Campoleoni:2011hg}, and
a 1-loop calculation \cite{Gaberdiel:2010ar} suggested that the corresponding 
statement would also be true for the quantum theory. A concrete proposal was
then made in \cite{Gaberdiel:2010pz}, relating the large $N$ 't~Hooft like limit
of the $W_N$ level $k$ minimal models 
to a family of bosonic higher spin theories on AdS$_3$. By now quite some evidence
has been found in favour of this proposal 
\cite{Gaberdiel:2011zw,Chang:2011mz,Papadodimas:2011pf,Ahn:2011by,Castro:2011iw,%
Ammon:2011ua,Chang:2011vk}.
The proposal is the natural analogue of the Klebanov-Polyakov duality 
since, for vanishing 't~Hooft coupling, the CFT can be 
described as the singlet sector of a free theory \cite{Gaberdiel:2011aa}.
There have also been interesting results concerning the construction of 
black holes for these higher spin theories, as well as their dual CFT interpretation 
\cite{Gutperle:2011kf,Ammon:2011nk,Kraus:2011ds,GHJ}. 
\smallskip

The proposal of  \cite{Gaberdiel:2010pz} was generalised to the case 
where instead of the $\mathfrak{su}(N)$ based $W$-algebras, one considers
the $\mathfrak{so}(2N)$ series \cite{Ahn:2011pv,Gaberdiel:2011nt}. More
recently, a ${\cal N}=2$ supersymmetric generalisation has been proposed
\cite{Creutzig:2011fe}, relating a family of Kazama-Suzuki models 
\cite{Kazama:1988qp,Kazama:1988uz} 
to the supersymmetric higher spin theory of  \cite{Prokushkin:1998bq,Prokushkin:1998vn}. 
It is the aim of this paper to give substantial evidence 
in favour of this proposal; in particular, we shall give the supersymmetric generalisation
of the calculation of \cite{Gaberdiel:2011zw}, establishing the agreement between
the 1-loop partition function of the supersymmetric higher spin theory on AdS$_3$, 
and the partition function of the dual ${\cal N}=2$ superconformal field theories in 
the large $N$ limit. While the general strategy is similar to what was done in  
\cite{Gaberdiel:2011zw}, there is one new ingredient in our analysis: unlike the bosonic 
$W_N$ case, explicit formulae for the coset
characters of the Kazama-Suzuki models do not appear to be readily available. 
In this paper we therefore calculate them from first principles in the 't~Hooft limit. The 
basic idea is to relate them to the branching functions of the free ($\lambda=0$) theory
which can be determined by combinatorial methods. We first apply this approach
to the bosonic case, thereby reproducing the results of \cite{Gaberdiel:2011zw}, 
and then use it for the supersymmetric Kazama-Suzuki models.
\medskip

The paper is organised as follows. In section~\ref{eq:non_susy_strory} we review
the bosonic duality; in particular, we explain in detail how the partition function of the 
minimal models can be calculated from first principles in the 't~Hooft limit, using
a combinatorial approach (see section~\ref{sec:ff_realization}). In section~\ref{sec:susy}
we then apply the same techniques to the 't~Hooft limit of the Kazama-Suzuki models.
Finally, section~4 contains our conclusions and an outlook towards future directions.
We have relegated some of the technical arguments for the calculation of the
branching and restriction rules for $\gl(\infty|\infty)_+$ (that play a role for the
supersymmetric analysis) to an appendix.



\section{Non-supersymmetric duality}
\label{eq:non_susy_strory}

In this section we briefly review the non-supersymmetric duality and rederive the
relation between the partition functions. Our strategy follows essentially \cite{Gaberdiel:2011zw}, 
but we employ a somewhat different technique for extracting explicit formulae for the
coset characters in the 't~Hooft limit. This method will generalise directly to
the supersymmetric case.

\subsection{The higher spin gravity theory}
\label{sec:hs_grav_theory}

Let us begin by fixing some conventions. We 
parametrise the Euclidean AdS$_3$ space with coordinates  
$(r,z)\in \mathbb{R}\times\mathbb{C}$, for which the metric takes the form
\begin{equation}
\mathrm{d}s^2=\frac{\mathrm{d}r^2+\mathrm{d}z\mathrm{d}\bar{z}}{r^2} \ ,
\end{equation}
and the boundary is located  at $r=0$. In thermal AdS the points 
$(r,z+\mathbb{Z}+\mathbb{Z}\tau)$ are identified, and the boundary becomes a 
torus with modular parameter $q=e^{2\pi i \tau}$. 
We shall first consider the non-supersymmetric truncation of Vasiliev's higher 
spin theory~\cite{Prokushkin:1998bq,Prokushkin:1998vn} on AdS$_3$. 
This theory has massless gauge fields of spin $s=2,3,\ldots$. Assuming periodic 
boundary conditions around  
the thermal circle, a real gauge field with integer spin $s$ contributes to the 1-loop
partition function the factor 
\begin{equation}\label{eq:b_gauge_pf}
Z_{\text{gauge}}^{s}=\prod_{n=s}^\infty \frac{1}{\vert1-q^n\vert^{2}}\ .
\end{equation}
This was first calculated for the graviton ($s=2$) in \cite{Giombi:2008vd}; the general result
was then derived in  \cite{Gaberdiel:2010ar} using the techniques of \cite{David:2009xg}.

In addition to these massless higher spin gauge fields, the theory that is proposed to be dual
to the 't~Hooft limit of the minimal model also contains two massive complex scalar fields
\cite{Gaberdiel:2010pz}. A  complex scalar field $\phi$ of mass squared $M^2$ 
contributes to the partition function the factor  \cite{Giombi:2008vd}
\begin{equation}\label{eq:scalar_pm}
 Z^\Delta_{\text{scalar}}=\prod_{m,n=0}^\infty\frac{1}{(1-q^{h+m}\bar{q}^{h+n})^2} \ ,
\end{equation}
provided its asymptotic behaviour near the AdS boundary is fixed to be 
$\phi(r,z,\bar{z})\sim a(z,\bar{z}) r^{\Delta}$. Here $\Delta=2h$ is related to the mass 
squared $M^2$ by the familiar relation
\begin{equation}
(\Delta -1 )^2 = 1 + M^2 \  .
\end{equation}
In the duality of \cite{Gaberdiel:2010pz} $M^2 = -1 + \lambda^2$, and then there are two
solutions for $\Delta$, 
\begin{equation}\label{eq:bos_scal_dim}
 \Delta_\pm^B(\lambda)=1\pm \lambda\    . 
\end{equation}
According to the proposal of \cite{Gaberdiel:2010pz}, one complex scalar is quantised 
with $(+)$ boundary conditions, the other with $(-)$ boundary conditions. Then the total 
1-loop partition function of the higher spin theory equals
\begin{equation}\label{eq:part_func_nonsusy}
 Z^\lambda_{\text{1-loop}} = Z^{\Delta^B_+(\lambda)}_{\text{scalar}}\times Z^{\Delta^B_-(\lambda)}_{\text{scalar}}\times \prod_{s=2}^\infty
Z^s_{\text{gauge}}\ .
\end{equation}
It was conjectured in \cite{Gaberdiel:2010pz} that this higher spin theory
is dual to a specific limit of minimal model CFTs that we shall now review.

\subsection{The coset point of view }\label{sec:formulation_nonsusy}

Consider the coset conformal field theory 
\begin{equation}\label{eq:sln_minimal}
 \frac{\su(N)_k\oplus \su(N)_1}{\su(N)_{k+1}}
\end{equation}
for integer level $k$. Its chiral algebra is the $W_N$ algebra of central charge
\begin{equation}
c=(N-1)\left(1-\frac{N(N+1)}{(N+k)(N+k+1)}\right)  \ ,
\end{equation}
which we denote as $W_{N,k}$. The primaries of the coset 
CFT~\eqref{eq:sln_minimal} can be described  in the usual manner \cite{GKO85,GKO86}.

In order to do so explicitly, let us introduce the following notation. We denote by 
$\Yan$ the set of all Young diagrams, and by $\Yan_N\subset\Yan$ the subset of diagrams
with less than $N$ rows; as is well known the elements of $\Yan_N$ label the  
representations of $\su(N)$. The representations of the affine algebra  $\su(N)_k$ 
at level $k$ are then described by the diagrams $\Yan_{N,k} \subset \Yan_N$ that 
have in addition less or equal than $k$ columns. 

For $\Lambda\in \Yan_{N,k}$ and 
$\omega\in \Yan_{N,1}$ consider the decomposition of the tensor product 
in terms of representations $\Xi\in\Yan_{N,k+1}$ of $\su(N)_{k+1}$
\begin{equation}\label{eq:primary_def}
\Lambda \otimes \omega = \bigoplus_{\Xi} 
(\Lambda;\Xi) \otimes \Xi  \ ,
\end{equation}
where $\su(N)_{k+1}$ is diagonally embedded into $\su(N)_{k}\oplus \su(N)_{1}$, and 
$(\Lambda;\Xi)$ denotes the corresponding multiplicity space.
It is clear that only those $\Xi\in \Yan_{N,k+1}$ can appear in \eqref{eq:primary_def}
for which the weights satisfy
\begin{equation}\label{selection}
\Lambda + \omega- \Xi\in Q_N\ ,
\end{equation}
where $Q_N$ is the root lattice of $\su(N)$. For $\su(N)$, this equation determines
$\omega$ uniquely in terms of $\Lambda$ and $\Xi$. The multiplicity 
spaces $(\Lambda;\Xi)$ can thus be labelled by just $\Lambda$ and $\Xi$,
and they carry, by construction, an action of the coset CFT~\eqref{eq:sln_minimal}. The coset 
CFT is rational and all its highest weight representations can be obtained in this manner;
however, not all pairs $(\Lambda;\Xi)$ define inequivalent coset representations,
since there are field identifications \cite{Gepner:1989jq, Lerche:1989uy,Moore:1989yh}.

Let us denote the characters of the $\su(N)_k$ and $W_{N,k}$ representations as 
\begin{equation}\label{eq:slnk_chars}
  \ch^{N,k}_\Lambda(q,e^H)=\tr_{\Lambda}q^{L_0}e^H\ , \qquad
   b^{N,k}_{\Lambda;\Xi}(q)=\tr_{(\Lambda;\Xi)} q^{L_0} \ .
\end{equation}
Here $L_0$ is the zero mode of the energy momentum tensor in the corresponding 
chiral algebra, while $H$ is an element of the Cartan subalgebra of $\su(N)$.
As a consequence of~\eqref{eq:primary_def}, we have the basic relation
\begin{equation}\label{eq:basic_char_rel_finite}
 \ch^{N,k}_\Lambda (q,e^H)\, 
 \ch^{N,1}_\omega(q,e^H) = \sum_{\Xi}  
 b^{N,k}_{\Lambda;\Xi}(q) \, \ch^{N,k+1}_\Xi(q,e^H) \ ,
\end{equation}
which we will use below in order to compute the characters of the coset theory.

The simplest coset CFT is the usual charge-conjugation theory, whose
Hil\-bert space consists of 
\begin{equation}\label{eq:sln_coset_space}
\mathcal{H}^{N,k} = \bigoplus_{[\Lambda;\Xi]} (\Lambda;\Xi) \otimes \overline{(\Lambda;\Xi)} \ ,
\end{equation}
where the two tensor factors are representations of the left- and right-moving coset CFT, 
respectively, and  the sum is taken over isomorphism classes $[\Lambda;\Xi]$ of 
representations identified by the field identification. 
The corresponding modular invariant torus partition function is then
%
%
%
\begin{equation}\label{eq:Zcoset}
  Z^{N,k}(q)=|q^{-\frac{c}{24}}|^2\sum_{[\Lambda;\Xi]} |b^{N,k}_{\Lambda;\Xi}(q)|^2\ .
\end{equation}

\begin{figure}[th]
\psfrag{a}{$\boldsymbol{\Lambda}_{N\to\infty}$}
\psfrag{b}{$\Lambda_l$}
\psfrag{c}{$\Lambda_r$}
\psfrag{l}{$N\to\infty$}
\centering{\includegraphics[scale=1]{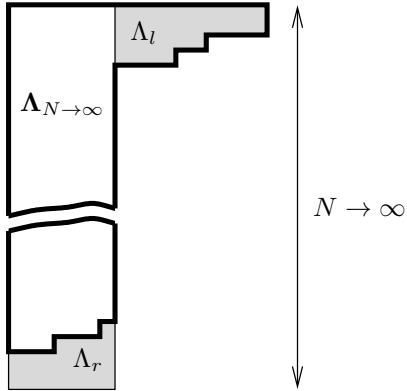}}
\caption{Young diagrams that are finite only in the horizontal direction, and 
that have  a single infinite vertical step label $\su(N)$ representations 
generated by the tensor product of finitely many fundamental and dual representations 
in the limit $N\to\infty$.
\label{fig:inf_young}
}
\end{figure}

It was proposed in~\cite{Gaberdiel:2010pz} that the non-supersymmetric higher spin theory of Vasiliev 
is dual to the 't~Hooft like large $N,k$ limit of the coset CFTs~\eqref{eq:sln_coset_space}, 
\begin{equation}\label{eq:t_hooft}
N,k\rightarrow \infty \qquad \hbox{with}\quad 
 \frac{N}{N+k}=\lambda \quad \hbox{held fixed.}
\end{equation}
A strong argument in favour of this proposal is the fact that the 
partition function~\eqref{eq:part_func_nonsusy} can be reproduced from the
dual CFT  in this limit. The way this happens is 
however quite intricate, since the naive limit of the partition function \eqref{eq:Zcoset}
diverges. In order to make sense of the limit theory it was proposed in \cite{Gaberdiel:2010pz} to 
restrict the Hilbert space~\eqref{eq:sln_coset_space} to those coset representations
for which both $\Lambda$ and $\Xi$ are contained in the  $N\rightarrow \infty$ limit 
of  finite tensor powers of the fundamental representation of $\su(N)$ and its dual.
Intuitively this means that in the limit both $\Lambda$ and $\Xi$ are described by 
a pair of Young diagrams, see fig.~\ref{fig:inf_young}.

\smallskip

\begin{figure}[t!]
\psfrag{1}{$1$}
\psfrag{2}{$2$}
\psfrag{3}{$3$}
\psfrag{a}{$-1$}
\psfrag{b}{$-2$}
\psfrag{0}{$0$}
\psfrag{d}{$\cdots$}
\psfrag{n}{$N$}
\psfrag{n1}{$N-1$}
\psfrag{lr}{$\Lambda_l$}
\psfrag{l}{$\boldsymbol{\Lambda}_N$}
\psfrag{ll}{$\Lambda_r$}
\centering{\includegraphics[scale=2]{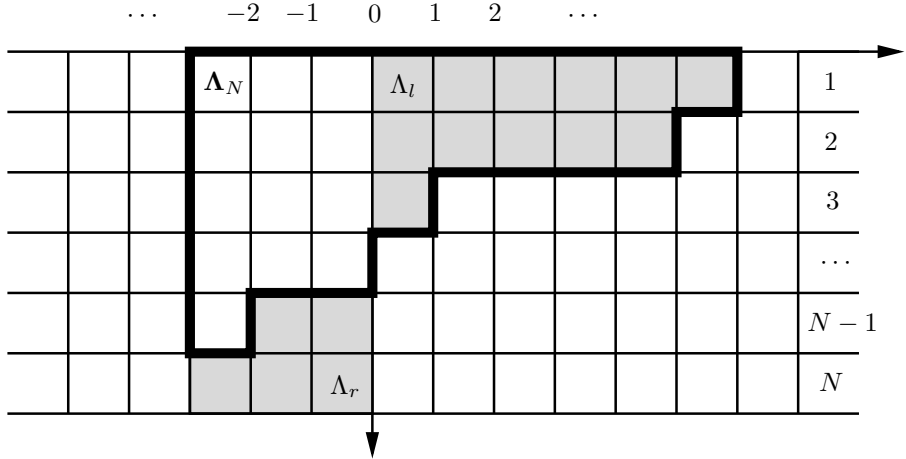}}
\caption{A $\un(N)$ representation is labelled by a
pair of (finite) Young diagrams $\boldsymbol{\Lambda}=(\Lambda_l,\Lambda_r)$ such that the sum of 
their rows is at most $N$. The corresponding $\su(N)$ dominant weight is represented 
by the Young diagram with a bold contour, denoted by $\boldsymbol{\Lambda}_N$.
\label{fig:gln_to_sln_diags}
}
\end{figure}

In order to explain this more precisely, it is convenient to think of these labels in terms
of $\un(N)$ representations. Recall that irreducible (tensorial) representations of $\un(N)$
are labelled by pairs of Young diagrams $\boldsymbol{\Lambda}=(\Lambda_l,\Lambda_r)$ of the form represented in fig.~\ref{fig:gln_to_sln_diags}.
Every $\un(N)$ tensor $\boldsymbol{\Lambda}$ defines an $\su(N)$
tensor labelled by a single Young diagram in $\Yan_N$, which we denote by  $\boldsymbol{\Lambda}_N=(\Lambda_l,\Lambda_r)_N$.
Since we can move the position where we separate ${\bf \Lambda}_N$ into $\Lambda_l$
and $\Lambda_r$,  there are 
many $\un(N)$ tensors $\boldsymbol{\Lambda}$ that restrict to the same $\su(N)$
tensor $\boldsymbol{\Lambda}_N$, but differ in their $\un(1)$ charge $|\boldsymbol{\Lambda}|_-=|\Lambda_r|-|\Lambda_l|$,
where $|\Lambda_{l,r}|$ is the number of boxes in the corresponding diagrams.

The representations we are interested in are those where we 
keep $\Lambda_l$ and $\Lambda_r$ fixed as we take the $N\rightarrow\infty$ limit;
the resulting $\su(N)$ representation  becomes then an infinite
%
%
Young diagram depicted in fig.~\ref{fig:inf_young}.
As can be seen from this figure, one can unambiguously recover back 
from this infinite Young diagram the original pair of finite Young diagrams 
$\boldsymbol{\Lambda}$.
From now on we shall identify the set of these infinite Young diagrams 
$\boldsymbol{\Lambda}_{\infty}=(\Lambda_l,\Lambda_r)_{\infty}$
with the set $\mathbf{Y}=Y\times Y$ of pairs of Young diagrams,
and denote its elements by bold upper case Greek letters (such as ${\bf \Lambda}$).

\smallskip

Returning to the limit of theory of (\ref{eq:sln_coset_space}), it 
was proposed in~\cite{Gaberdiel:2010pz} that the Hilbert space 
reproducing the partition function of the dual AdS$_3$ theory 
in the limit~\eqref{eq:t_hooft} is
%
\begin{equation}\label{eq:finite_fusion_states}
{\cal H}^\lambda =
 \bigoplus_{\boldsymbol{\Lambda},\boldsymbol{\Xi}\in \mathbf{Y}}
(\boldsymbol{\Lambda}; \boldsymbol{\Xi}) \otimes
\overline{(\boldsymbol{\Lambda} ;\boldsymbol{\Xi})}\ ,
\end{equation}
where the two pairs of Young diagrams $\boldsymbol{\Lambda}=(\Lambda_l,\Lambda_r)$ and $\boldsymbol{\Xi}=(\Xi_l,\Xi_r)$
label representations
\begin{equation}\label{eq:rep_winf}
(\boldsymbol{\Lambda};\, \boldsymbol{\Xi})=\lim_{N,k\to\infty} \bigl( \boldsymbol{\Lambda}_N ;\boldsymbol{\Xi}_N \bigr)
\end{equation}
of the limit algebra
$W_{\infty}[\lambda]$ \cite{Gaberdiel:2011wb}. 


The second complication comes from the fact that the representations~\eqref{eq:rep_winf}
generically become reducible in the limit \eqref{eq:t_hooft}, 
at least if both $\boldsymbol{\Lambda}$ and $\boldsymbol{\Xi}$ are non-trivial.
Another way of saying this is that 
new null states appear in the limit that have to be removed in order to calculate
the partition function. Subtracting out these contributions, it was 
argued in \cite{Gaberdiel:2011zw} that the resulting partition
function of \eqref{eq:finite_fusion_states} reproduces precisely 
\eqref{eq:part_func_nonsusy}.

We would now like to give a modified version of the proof and then generalise it to the 
supersymmetric case.

\subsection{The character identity}
\label{sec:proof}

In order to make contact with  eq.~(\ref{eq:finite_fusion_states}),
the first step of the argument is to rewrite the bulk partition 
function~\eqref{eq:part_func_nonsusy} as a sum over 
finite Young diagrams. Let us begin by introducing a little bit of notation. Let
$\gl(\infty)_+$ be the Lie algebra of infinite-dimensional matrices 
for which only finitely many diagonals adjacent to the main diagonal are non-zero. These matrices have a natural
action on the infinite-dimensional vector space 
$\mathbb{C}^{\mathbb{N}_0} =\oplus_{j=0}^\infty \mathbb{C}e_j$, where 
$e_j$ denotes a basis and $\mathbb{N}_0$ are the non-negative integers. 
This is the `fundamental' representation of $\gl(\infty)_+$, and the representations
we are interested in are those that are contained in finite tensor powers of this
fundamental representation. All of these tensor products are completely
decomposable, and hence we can label these representations by finite Young 
diagrams $\Lambda$.

We shall need to calculate the character  of the representation $\Lambda$.
In general, a character can be evaluated on an arbitrary element of the 
Cartan subgroup of the associated group $\GL(\infty)_+$. The Cartan subgroup
consists of the diagonal matrices, and 
the Cartan subalgebra of $\gl(\infty)_+$ can thus also be identified with the diagonal 
matrices; a natural basis for the Cartan subalgebra is $H_i=E_{ii}$, where
$i\in {\mathbb N}_0$ and 
$E_{ii}$ is the matrix with a single non-zero entry in position $(i,i)$. The dual
to the Cartan subalgebra is the weight space, and it is generated by the weights
$\epsilon_i$ with
\begin{equation}\label{fundweight}
\epsilon_i (H_j) = \delta_{ij} \ .
\end{equation}

With these preparations we can now describe the character of the representation
$\Lambda$. A basis for the vector space associated to $\Lambda$ 
is labelled by the different Young tableaux ${\rm Tab}_\Lambda$ 
of shape $\Lambda$. Here a Young tableaux of shape $\Lambda$ 
is a Young diagram $\Lambda$ 
together with a filling of the boxes of $\Lambda$ by elements from $\mathbb{N}_0$,
where, as usual, within each row the entries of the boxes do not decrease, while within 
each column they increase. The weight $\wt(T)$  of the basis element associated to 
$T\in {\rm Tab}_\Lambda$ is the sum of the associated weights $\epsilon_j$, 
where $j$ runs over the entries of the boxes in the tableau $T$. Then the character of 
$\Lambda$ equals
\begin{equation}\label{eq:def_char_comb}
  \ch_\Lambda (e^H) = \sum_{T\in \Tab_\Lambda} e^{\wt(T)(H)}\ ,
\end{equation}
where $H$ is an arbitrary element of the Cartan subalgebra of of $\gl(\infty)_+$.
In the following we shall mainly evaluate this character on the specific
elements 
\begin{equation}
U(h)=\prod_{j\in \mathbb{N_0}} e^{ 2\pi i \tau(h+j)H_j}  \ , \qquad q = e^{2\pi i \tau} 
\end{equation}
of the Cartan subgroup of $\GL(\infty)_+$ with matrix elements 
\begin{equation}
U(h)_{jj} = q^{h+j} \ ,
\end{equation}
where $h$ is some real number, and $q$ has modulus less than one. 
In this case the  character~\eqref{eq:def_char_comb} takes the form
\begin{equation}\label{eq:ev_ch}
\ch_\Lambda (U(h))=\sum_{T\in\Tab_\Lambda}\prod_{j\in T}q^{h+j}\ .
\end{equation}

With the help of the matrix $U(h)$, we can now write  the partition function of  
a real scalar field on thermal AdS$_3$  as the determinant
\begin{equation}\label{eq:det_uinf}
\prod_{m,n=0}^\infty\frac{1}{1-q^{h+m}\bar{q}^{h+n}} 
= \frac{1}{\det (1- U(h)\otimes U(h)^\ast)} \ ,
\end{equation}
where $U(h)^\ast$ is obtained from $U(h)$ upon replacing $q\mapsto \bar{q}$.
This can be decomposed into $\gl(\infty)_+$ characters 
by performing the same  manipulations as 
in~\cite{Gaberdiel:2011zw}\footnote{This expansion formally defines the Schur 
functions (with an infinite number of variables) in the theory of symmetric functions 
\cite{macdonald}. Their explicit expression as a sum over monomials labelled by 
Young tableaux of fixed shape and, thus the identification with $\gl(\infty)_+$ characters, 
is then an {\em a posteriori} fact.}
\begin{equation}\label{eq:comb_id_1}
  \frac{1}{\det (1- U(h)\otimes U(h)^*)} = 
  \sum_{\Lambda} \ch_{\Lambda}(U(h))\, \overline{\ch_{\Lambda} (U(h)) }\ .
\end{equation}
The partition function~\eqref{eq:part_func_nonsusy} can thus be written as
\begin{equation}\label{eq:comparable_form}
  Z_{\text{1-loop}}^\lambda= Z_{\text{gauge}}\sum_{\Lambda_l,\Lambda_r,\Xi_l,\Xi_r}
|\ch_{\Lambda_l}(U_+)\, \ch_{\Lambda_r}(U_+)\, \ch_{\Xi_l}(U_-)\, \ch_{\Xi_r}(U_-)|^2\ ,
\end{equation}
where $\Lambda_l,\Lambda_r,\Xi_l,\Xi_r$ are finite Young diagrams, and
we have defined
\begin{equation}
 Z_{\text{gauge}}=\prod_{s=2}^\infty Z_{\text{gauge}}^s\ ,
\end{equation}
with $Z_{\text{gauge}}^s$ given in \eqref{eq:b_gauge_pf}.
Finally, $U_\pm\equiv U(h_\pm)$, with $h_\pm = \tfrac{1}{2}(1\pm \lambda)$. 
\medskip

The next step is to reproduce eq.~\eqref{eq:comparable_form} from the coset
point of view. To this end we need to evaluate the coset characters 
up to powers of $q^k$ or $q^N$, which become irrelevant in the 't~Hooft limit. 
We want to determine the coset characters from \eqref{eq:basic_char_rel_finite},
and thus we first need to understand the characters of affine representations
associated to $\Lambda\in\Yan_{N,k}$. It follows from the Kac-Weyl formula
(see e.g.\ \cite{BouScho93,Gaberdiel:2011zw}) that we have 
\begin{equation}\label{eq:limit_affine_char}
 \ch^{N,k}_{\Lambda}(q,e^H)= \frac{q^{h^{N,k}_{\Lambda}}
[\ch^N_{\Lambda}(e^H)+\mathcal{O}(q^{k-\Lambda_1+1})]}
{\prod_{n=1}^\infty\left[(1-q^n)^{N-1}\,
\prod_{\alpha\in\Delta_N}(1-q^ne^{\alpha(H)})\right]}\ ,
\end{equation}
where $\Delta_N$ denotes the roots of $\su(N)$,
 and $\Lambda_1$ is the length of the first row of $\Lambda$.
Here we have used that, for large $k$, 
only the elements of the finite Weyl group contribute to the dominant term,
thus making the finite $\su(N)$ character $\ch^N_{\Lambda}$ appear.
The conformal dimension of the affine primary field labelled by $\Lambda$ equals
\begin{equation}\label{eq:conf_dim_aff_primary}
 h^{N,k}_\Lambda = \frac{\Cas(\Lambda) }{2(k+N)}=
\frac{N|\Lambda|}{2(k+N)}-\frac{|\Lambda|^2}{2N(k+N)}+\sum_{\varepsilon\in\Lambda}\frac{\col(\varepsilon)-\row(\varepsilon)}{k+N}\ ,
\end{equation}
where $|\Lambda|$ is the number of boxes in $\Lambda$, and
the sum in the last term runs over  the individual boxes of the Young diagram $\Lambda$, 
 where $\row (\varepsilon)$ and $\col(\varepsilon)$
is the row and column number of the box $\varepsilon\in\Lambda$, respectively.

\noindent Using (\ref{eq:limit_affine_char}) for the different characters in \eqref{eq:basic_char_rel_finite}
we thus obtain
\begin{equation}\label{eq:starting_def_aux_branch}
 \ch^N_{\Lambda}(e^H)\, \ch_{\omega}^{N,1}(q,e^H)=
 \sum_{\Xi\in Y_N}\, a^{N}_{\Lambda;\Xi}(q) \, \ch^N_{\Xi}(e^H)\ ,
\end{equation}
where we have defined the $k$ independent function $a^{N}_{\Lambda;\Xi}(q)$; 
it is related to the coset character in the $k\to\infty$ limit  as 
\begin{equation}\label{eq:branching_lambda_ind}
 b^{N,k}_{\Lambda;\Xi}(q)= q^{h^{N,k}_{\Lambda}-\,h^{N,k+1}_{\Xi}} \, 
 \Big[a^{N}_{\Lambda;\Xi}(q)+\mathcal{O}(q^{k-\Lambda_1+1})+\mathcal{O}(q^{k-\Xi_1+2})\Big] \ .
\end{equation}
Next
we observe that for $\Lambda=0$ eq.~\eqref{eq:starting_def_aux_branch} simplifies to 
\begin{equation}\label{eq:def_elem_branching_func_nonsusy}
 \ch^{N,1}_{\omega}(q,e^H) = \sum_{ \Xi\in Y_N}\,
 a^{N}_{0;\Xi}(q)\, \ch^N_{\Xi}(e^H) \ ,
\end{equation}
i.e.\  $a^{N}_{0;\Xi}(q)$ is the branching function of an $\su(N)_1$ affine 
representation into representations of the zero mode algebra $\su(N)$. 
%
%
In order to describe the general case, recall that the decomposition of $\su(N)$ tensor products
implies that
\begin{equation}
\ch^N_{\Lambda_1} \ch^N_{\Lambda_2} = \sum_{\Lambda_3\in Y_N} 
c_{\Lambda_1\Lambda_2}^{(N)\, \Lambda_3}\, \ch^N_{\Lambda_3} \ ,
\end{equation}
where $c_{\Lambda_1\Lambda_2}^{(N)\, \Lambda_3}$ are the 
Clebsch-Gordan coefficients. Multiplying 
\eqref{eq:def_elem_branching_func_nonsusy} by  $\ch^N_{\Lambda}$ we 
thus conclude that 
\begin{equation}\label{eq:gen_branch_func_explicit}
a_{\Lambda;\Xi}^{N}(q)=
\sum_{\Pi\in Y_N}c_{\Lambda \Pi}^{(N)\, \Xi}\,  a^{N}_{0;\Pi}(q) 
= \sum_{\Pi\in Y_N}c_{\Lambda \bar{\Xi}}^{(N)\, \bar{\Pi}}\,  a^{N}_{0;\Pi}(q)  \ ,
\end{equation}
where the bar denotes the conjugate representation and for the second equality we have used the symmetries of the 
Clebsch-Gordan coefficients.
%
%

Up to now all the equations are valid for finite $k$ and finite $N$.
As we have mentioned in sec~\ref{sec:hs_grav_theory}, in the large  
$N,k$ limit~\eqref{eq:t_hooft}
we shall restrict $\Lambda$ and $\Xi$ to be a those special infinite Young diagrams 
that can be identified with 
pairs of finite Young diagrams, see fig.~\ref{fig:inf_young}.
Using a free fermion construction, we will show in 
section~\ref{sec:ff_realization} that the power series expansion of $a^N_{0;\Xi}(q)$  
stabilises in the large $N$ limit to
\begin{equation}  \label{eq:FF_branchings}
\begin{array}{rcl}
a_{0;0}(q)  &= & 
{\displaystyle \lim_{N\to\infty}a_{0;0}^N(q)
=\prod_{s=2}^\infty\prod_{n=s}^\infty\frac{1}{1-q^n}} \\[10pt]
a_{0;\boldsymbol{\Xi}}(q)
& =  &  
{\displaystyle \lim_{N\to\infty}a^N_{0;\boldsymbol{\Xi}_N}(q)
 =\ch_{\Xi^t_l}(U_0)\, \ch_{\Xi^t_r}(U_0) \, a_{0;0}(q) \ , }
\end{array}
\end{equation}
where $\boldsymbol{\Xi}=(\Xi_l,\Xi_r)$ and $U_0=U\big(h=\tfrac{1}{2}\big)$.  Note 
that $|a_{0;0}(q)|^2 =Z_{\rm gauge}(q)$. It follows by a direct calculation 
(see e.g.\ \cite[eq.\ (2.7)]{Gross:1993hu}) that
\begin{equation}
\Cas(\boldsymbol{\Lambda}_N)=\Cas(\Lambda_l)+\Cas(\Lambda_r)
+\frac{2|\Lambda_l||\Lambda_r|}{N}\ .
\end{equation}
Thus,  for large $N$ the conformal dimensions of  the affine primaries behave as
\begin{equation}\label{eq:conf_dim_aff_primary_ass}
 h^{N,k}_{\boldsymbol{\Lambda}_N} = 
 \frac{N \Bigl( |\Lambda_l|+|\Lambda_r|\Bigr)}{2 (k+N)} + {\cal O}\left( \tfrac{1}{N} \right)  \ ,
\end{equation}
where $\boldsymbol{\Lambda}=(\Lambda_l,\Lambda_r)$ is a pair of finite Young diagrams.
Hence the exponent of the prefactor in eq.~\eqref{eq:branching_lambda_ind} becomes 
in the 't~Hooft limit 
\begin{equation}
 \lim_{k,N\to\infty} \Bigl( h^{N,k}_{\boldsymbol{\Lambda}_N}  -  h^{N,k+1}_{\boldsymbol{\Xi}_N} \Bigr)
 = \frac{\lambda}{2}\, \bigl( |\boldsymbol{\Lambda}| - |\boldsymbol{\Xi}|\bigr) \ ,
\end{equation}
where we have defined  $|{\bf \Lambda}|=|\Lambda_l|+|\Lambda_r|$ and similarly  $|{\bf \Xi}|$.
Thus, the branching functions corresponding to the $W_\infty[\lambda]$ modules~\eqref{eq:rep_winf}
have the explicit form
\begin{equation}\label{eq:b_lim_expl}
b^\lambda_{\boldsymbol{\Lambda};\boldsymbol{\Xi}}(q)=
q^{\frac{\lambda}{2}(|\boldsymbol{\Lambda}|-\boldsymbol{|\Xi|})} a_{0;0}(q)
\sum_{\boldsymbol{\Pi}\in \mathbf{Y}}
c_{\boldsymbol{\Lambda}\bar{\boldsymbol{\Xi}}}^{\ \; \bar{\boldsymbol{\Pi}}}\,
\ch_{\Pi^t_l}(U_0)\ch_{\Pi^t_r}(U_0)\ ,
\end{equation}
where for  $\boldsymbol{\Pi}=(\Pi_l,\Pi_r)$ the conjugate representations is 
$\bar{\boldsymbol{\Pi}}=(\Pi_r,\Pi_l)$.

It was argued in \cite{Gaberdiel:2011zw} that subtracting out the null-states (see the discussion at the
end of section~\ref{sec:formulation_nonsusy}) is equivalent to restricting ${\bf \Pi}$ in 
eq.~\eqref{eq:b_lim_expl} to those Young diagrams that satisfy
\begin{equation}\label{eq:constraint}
 |{\bf \Lambda}|+|{\bf \Xi}|= |{\bf \Pi}|\ .
\end{equation}
Note that this is similar to what happens for the tensor product decomposition for 
$\su(N)$ in the large $N$ limit. For example, in the tensor product of the fundamental and
anti-fundamental representation of $\su(N)$, the projector onto the $\su(N)$ invariant state
is of the form 
\begin{equation}\label{eq:singlet}
\sum_{j=1}^{N} \frac{1}{N} e_j \otimes e^j \  ,
\end{equation}
where $e_j$ and $e^j$ are a basis and the dual basis for the fundamental and 
anti-fundamental representation, respectively. In the large $N$ limit \eqref{eq:singlet}
vanishes, and the tensor product is no longer completely decomposable. In our case,
the analogue of (\ref{eq:singlet}) are the states where $|{\bf \Lambda}| + |{\bf \Xi}| <|{\bf \Pi}|$,
and they vanish in the large $N$ limit as demonstrated (in some simple examples) in 
\cite{Gaberdiel:2011zw}. In terms of the Clebsch-Gordan coefficients, 
\eqref{eq:constraint}
implies that 
\begin{equation}
c_{\boldsymbol{\Lambda}\bar{\boldsymbol{\Xi}}}^{\  \; \bar{\boldsymbol{\Pi}}}=
\lim_{N\to\infty} c
\begin{smallmatrix}  (N)&  \bar{\boldsymbol{\Pi}}_N\\  \boldsymbol{\Lambda}_N &\bar{\boldsymbol{\Xi}}_N \end{smallmatrix}
=
c_{\Lambda_l\Xi_r}^{\ \; \; \Pi_r} \, c_{\Lambda_r\Xi_l}^{\ \ \; \Pi_l}\ .
\end{equation}
We have furthermore used that the Clebsch-Gordan 
coefficients on the right hand side stabilise in the large $N$ limit.\footnote{In the theory of 
symmetric functions the numbers $c_{\Lambda\Xi}^{\; \ \Pi}$ are known as the 
Littlewood-Richardson coefficients, see~\cite[ch.~1]{macdonald}. Essentially, these {\em are} 
the Clebsch-Gordan coefficients of $\gl(\infty)_+$.}
Putting  everything together we then obtain for the trace over ${\cal H}^\lambda$ 
\begin{align} \notag
{\rm Tr}_{\mathcal{H}^\lambda}q^{L_0}\bar{q}^{\bar{L}_0}=
&Z_{\text{gauge}}\sum_{{\bf \Lambda},{\bf \Xi}}\Big|
q^{\frac{\lambda}{2}(|{\bf \Lambda}|-|{\bf \Xi}|)}
\sum_{\Pi_l,\Pi_r} c_{\Lambda_l\Xi_r}^{\; \; \ \Pi_r}\,
c_{\Lambda_r\Xi_l}^{\; \ \ \Pi_l}\,
\ch_{\Pi^t_l}(U_0) \, \ch_{\Pi^t_r}(U_0)
\Big|^2\\ \notag
=& Z_{\text{gauge}}\sum_{{\bf \Lambda},{\bf \Xi}}  
\Big|
q^{\frac{\lambda}{2}(|{\bf \Lambda}|-|{\bf \Xi}|)}
\ch_{\Lambda^t_l}(U_0)\, \ch_{\Xi^t_r}(U_0)\,
\ch_{\Lambda^t_r}(U_0)\, \ch_{\Xi^t_l}(U_0) \Big|^2\\
=&
Z_{\text{gauge}}\mathop{\sum_{\Lambda_l,\Lambda_r}}_{\Xi_l,\Xi_r}\Big|
\ch_{\Lambda^t_l}(U_+)\, \ch_{\Xi^t_r}(U_-)\, 
\ch_{\Lambda^t_r}(U_+)\, \ch_{\Xi^t_l}(U_-) \Big|^2  \ ,
\end{align}
where we have used that the Clebsch-Gordan coefficients are invariant under taking
transposes. This then agrees with \eqref{eq:comparable_form}.
%

\subsection{Free field realisation}\label{sec:ff_realization}

Finally we come to the proof of the two fundamental eqs.~\eqref{eq:FF_branchings};
this is where our analysis differs from \cite{Gaberdiel:2011zw}.  Recall that 
we can realise $\su(N)_1\oplus {\mathfrak u}(1)_N$ in terms of 
$N$ free Dirac fermions. Here ${\mathfrak u}(1)_N$ is the chiral ${\mathfrak u}(1)$ 
algebra $[J_m,J_n]=N\delta_{m,-n}$ 
that is extended by two fields of conformal dimension $h=\tfrac{N}{2}$ and
${\mathfrak u}(1)$-charge $\pm N$, see \cite[sec.~14.4.4]{yellow}. (Incidentally, ${\mathfrak u}(1)_N$ is
also the chiral algebra of a compact boson compactified at $R = \sqrt{N}$ where $R=\sqrt{2}$
describes the self-dual radius, i.e.\  ${\mathfrak u}(1)_2 \cong \mathfrak{su}(2)_1$.)
The irreducible representations of ${\mathfrak u}(1)_N$ are labelled by $l\in {\mathbb Z}_N$,
and their  characters are 
\begin{equation}\label{eq:u1_affine_chars}
  \Theta^N_l(q,w)= {\rm Tr}_l \bigl( q^{L_0} w^{J_0} \bigr) = 
  \sum_{m\in\mathbb{Z}}w^{l+Nm}\, 
  \frac{q^{\frac{1}{2N}(l+Nm)^2}}{\prod_{n=1}^\infty(1-q^n)}\ .
\end{equation}
On the level of characters, the relation between the free fermion theory and 
$\su(N)_1\oplus {\mathfrak u}(1)_N$ amounts then to 
\begin{equation}\label{eq:ff_sln_bos_branch}
 \prod_{n=0}^\infty \prod_{i=1}^N(1+wv_i q^{n+\frac{1}{2}}) \,
 (1+\bar{w}\bar{v}_i q^{n+\frac{1}{2}}) =
\sum_{l=0}^{N-1}\Theta_l^N(q,w)\,  \ch^{N,1}_{\omega_l}(q,v)\ ,
\end{equation}
where $\omega_l$  is the $l$-th fundamental weight of $\su(N)$, $w\in\U(1)$ and $v_i$ are 
the diagonal entries of an element $v$ of the Cartan torus of ${\rm SU}(N)$.
In terms of the branching functions introduced in 
eq.~\eqref{eq:def_elem_branching_func_nonsusy}, this then becomes 
\begin{equation}\label{eq:before_limit_ff_branch}
\prod_{n=0}^\infty \prod_{i=1}^N(1+wv_i q^{n+\frac{1}{2}}) \,
(1+\bar{w}\bar{v}_i q^{n+\frac{1}{2}}) =
\sum_{\Lambda\in Y_N} \Theta^N_{[\Lambda]}(q,w)\, 
a^{N}_{0,\Lambda}(q)\, \ch^N_\Lambda(v)\ ,
\end{equation}
where $[\Lambda]\in\mathbb{Z}_N$ denotes the congruence class of an 
$\su(N)$ representation $\Lambda$, see~\cite[sec.~13.1.9]{yellow}. For the following it 
is more convenient to decompose this partition function into characters of $\un(N)$ 
rather then $\su(N)$, i.e.\ to absorb the $w$-dependent factor of $\Theta^N_{[\Lambda]}$
into the $\un(N)$ character as 
\begin{equation}\label{eq:glN_to_slN_ch}
\ch^N_{\boldsymbol{\Lambda}_N}(v)\, w^{|\Lambda_r|-|\Lambda_l|} = 
\ch^N_{\boldsymbol{\Lambda}}(vw) \ .
\end{equation}
Recall that $\un(N)$ representations are parametrised by pairs of 
Young diagrams $\boldsymbol{\Lambda}=(\Lambda_l,\Lambda_r)$, see 
fig.~\ref{fig:gln_to_sln_diags}; alternatively, we may label them by a single
Young diagram $\boldsymbol{\Lambda}_N$ together with an integer $|\boldsymbol{\Lambda}|_-=|\Lambda_r| - |\Lambda_l|$ determining the $\un(1)$-charge of the
representation.
Using \eqref{eq:glN_to_slN_ch}, we can now rewrite \eqref{eq:before_limit_ff_branch} as
\begin{equation}\label{eq:ffid}
\prod_{n=0}^\infty \prod_{i=1}^N(1+wv_i q^{n+\frac{1}{2}}) \,
(1+\bar{w}\bar{v}_i q^{n+\frac{1}{2}}) = 
\sum_{\boldsymbol{\Lambda}\in\mathbf{Y}} d_{\boldsymbol{\Lambda}}^N(q) \, 
\ch^N_{\boldsymbol{\Lambda}}(vw)\ ,
\end{equation}
where 
\begin{equation}\label{eq:central_object_sec_FF}
d_{\boldsymbol{\Lambda}}^N(q)=
\frac{q^{\frac{1}{2N}(|\Lambda_r|-|\Lambda_l|)^2}} {\prod_{n=1}^{\infty}(1-q^n)}\, 
a_{0,\boldsymbol{\Lambda}_N}^{N}(q) 
\end{equation}
counts the number of $\un(N)$ tensors  
$\boldsymbol{\Lambda}$ which appear in the free fermion theory. In the following 
we shall compute~\eqref{eq:central_object_sec_FF} combinatorially. Note that
the prefactor in the numerator will become irrelevant for $N\rightarrow \infty$. 

Let us denote by $\psi^1,\ldots,\psi^N$ the $N$ Dirac fermions, with
$\bar\psi^1,\ldots,\bar\psi^N$ their complex conjugates. The vector space whose
character is the left-hand-side of (\ref{eq:ffid}) is spanned by the vectors of the form
\begin{equation}\label{eq:ffFock}
\prod_{j=1}^{n_{\bar{\psi}}} \bar\psi^{a_j}_{-r_j-\frac{1}{2}} \, \prod_{k=1}^{n_\psi} \psi^{b_k}_{-s_k-\frac{1}{2}}  \Omega \ ,
\end{equation}
where $a_j,b_k\in\{1,\ldots, N\}$, $r_j,s_k\in \mathbb{N}_0$, and $\Omega$ is the vacuum. 
These states fall into representations of the two commuting Lie algebras: $\un(N)$ 
acting on the indices $a_j$ and $b_k$; and $\gl(\infty)_+$ acting on the 
mode numbers $r_j,s_k$.\footnote{Note that 
the action of $\gl(\infty)_+$ on the modes of $\psi^a$, and on the modes of $\bar{\psi}^a$
is in both cases the fundamental representation of $\gl(\infty)_+$.}
The branching function $d^N_{\boldsymbol{\Lambda}}$ 
in \eqref{eq:ffid} counts the multiplicity with which the $\un(N)$ representation 
$\boldsymbol{\Lambda}$ appears in the Fock space, and because of the 
commuting $\gl(\infty)_+$ action, it will naturally be a character of $\gl(\infty)_+$.

More precisely, a $\un(N)$ tensor of shape $\boldsymbol{\Lambda}$ appears `for
the first time' (i.e.\ with multiplicity at most one) 
in the states of the form (\ref{eq:ffFock}) if $n_{\bar{\psi}}=|\Lambda_l|$ and $n_{\psi}=|\Lambda_r|$.
For a given choice of mode numbers $r_j$ and $s_k$, 
the multiplicity is precisely one if the $\{r_j\}$ and $\{s_k\}$ define an allowed filling of the
Young diagram $\Lambda_l^t$ and $\Lambda_r^t$, respectively, where 
$\Lambda^t$ denotes the transposed Young diagram -- this just keeps track of the
fact that, because of Fermi-Dirac statistics, the product of two identical fermionic modes
vanishes. If we sum over all such mode numbers (while keeping 
$n_{\bar{\psi}}=|\Lambda_l|$ and $n_{\psi}=|\Lambda_r|$ fixed), 
it follows from \eqref{eq:ev_ch}  that the branching function equals
\begin{equation}\label{eq:mult1}
\ch_{\Lambda_l^t}(U_0)\, \ch_{\Lambda_r^t}(U_0)\ ,
\end{equation}
where $U_0=U(h=\frac{1}{2})$. 

In order to complete the argument we only need to count the multiplicities with which
the $\un(N)$ representations $\boldsymbol{\Lambda}$ appear. As 
we have explained above, a given $\boldsymbol{\Lambda}$ 
appears `for the first time' if $n_{\bar{\psi}}=|\Lambda_l|$ and $n_{\psi}=|\Lambda_r|$.
However, it will continue to appear if $n_{\bar{\psi}}=|\Lambda_l|+m$ and 
$n_{\psi}=|\Lambda_r|+m$ with $m\in\mathbb{N}$, i.e.\ the state can be a product of 
a state with minimal number of factors, times a $\un(N)$ invariant state. Thus we need to count
also the $\un(N)$ invariants; according to the first fundamental theorem of classical 
invariant theory (see e.g.\ \cite{weyl}), all $\un(N)$ invariant states are linear combinations of the `basic' ones
\begin{equation}\label{eq:general_invariant}
  \prod_{r,s=0}^\infty 
\left( \sum_a \bar\psi^a_{-r-\frac{1}{2}} \psi^a_{-s-\frac{1}{2}} \right)^{M_{rs}}\ ,
\end{equation}
where only finitely many multiplicities $M_{rs}$ are non-zero. Note that not all of these
states are non-trivial; indeed, (\ref{eq:general_invariant}) vanishes if 
\begin{equation}
\sum_r M_{rs} > N \qquad \hbox{or} \qquad \sum_s M_{rs} > N \ .
\end{equation}
Furthermore, the states corresponding to different choices of $\{M_{rs}\}$ are not
all linearly independent; for example, for $N=1$ the two states whose non-zero multiplicities
are $\{M_{00}=1, M_{11}=1 \}$ and $\{M_{01}=1, M_{10}=1\}$ are in fact linearly dependent.
If we ignore these issues we can easily count the invariant tensors as 
\begin{equation}
d_{0}(q)  =  \prod_{r,s=0}^\infty \, \sum_{M_{rs}=0}^{\infty} q^{(r+s+1) M_{rs}} 
= \prod_{r,s=0}^{\infty} \frac{1}{1-q^{r+s+1}}  
=  \prod_{s=1}^{\infty} \prod_{n=s}^{\infty} \frac{1}{1-q^n} \ .
\end{equation}
This result is exact in the $N\to\infty$ limit because for finite $N$ the overcounting starts at 
order $q^{N+1}$ with the state
\begin{equation}
\left(\sum_a \bar{\psi}^a_{-\frac{1}{2}}\psi^a_{-\frac{1}{2}}\right)^{N+1}\ .
\end{equation}
Combining this result with (\ref{eq:mult1}), and ignoring the subtlety that 
the product of a non-vanishing scalar and a state of the form \eqref{eq:ffFock}
in some representation  $\boldsymbol{\Lambda}$ can also vanish --- again this
can be ignored in the $N\to\infty$ limit ---  then leads precisely to 
\eqref{eq:FF_branchings}. (Recall that the relation between 
$a_{0,\boldsymbol{\Lambda}}^N$ and $d_{\boldsymbol{\Lambda}}^N$ is given in 
\eqref{eq:central_object_sec_FF}.)



\section{Supersymmetric duality}\label{sec:susy}

In the following we want to generalise the above argument to the supersymmetric
setting. We begin by reviewing the structure of the supersymmetric higher spin theory.

\subsection{Higher spin supergravity}
\label{sec:hs_sugra}

The $\mathcal{N}=2$ supersymmetric higher spin 
supergravity theory of Prokushkin and Vasiliev~\cite{Prokushkin:1998bq, Prokushkin:1998vn}
has two (real) bosonic gauge fields of each spin $s=2,3,\ldots$, 
together with a single  current of spin $s=1$. In addition there are two (real) fermionic gauge 
fields for each spin $s=\tfrac{3}{2},\tfrac{5}{2},\ldots$. As in the bosonic case above, 
the structure of the theory depends on a real parameter $0\leq \lambda \leq 1$. However,
this parameter does not affect the quadratic part of the action, and the total 1-loop contribution
of the gauge fields equals
\begin{equation}
\mathcal{Z}_{\text{gauge}} = Z^1_{\text{gauge}}
\prod_{s=2}^\infty \left(Z^{s}_{\text{gauge}} Z^{s-\frac{1}{2}}_{\text{gaugino}}\right)^2 \ , 
\end{equation}
where the contribution of a real gauge field of half-integer spin $s$ 
\begin{equation}\label{eq:f_gauge_pf}
 Z_{\text{gaugino}}^{s}=\prod_{n=s-\frac{1}{2}}^\infty \vert 1+q^{n+\frac{1}{2}} \vert^{2}
\end{equation}
was calculated in  \cite{Creutzig:2011fe}.
We have assumed here that these half-integer spin gauge fields have 
anti-periodic boundary conditions around the thermal circle; from the dual CFT point of 
view, we shall therefore only consider the NS-sector. We also note that we can 
write $\mathcal{Z}_{\text{gauge}}$ in a manifestly supersymmetric way as
\begin{equation}\label{gauge1}
\mathcal{Z}_{\text{gauge}} = \prod_{s=1}^{\infty} \mathcal{Z}^s_{\text{gauge}}\ , \qquad
\hbox{where} \qquad
 \mathcal{Z}_{\text{gauge}}^s = Z_{\text{gauge}}^{s} 
 \left(Z_{\text{gauge}}^{s+\frac{1}{2}}\right)^2 Z_{\text{gauge}}^{s+1} \ ,
\end{equation}
with $\mathcal{Z}_{\text{gauge}}^s$ the contribution of the ${\cal N}=2$ gauge multiplet 
of integer spin $s$.

While the parameter $\lambda$ from above does not appear in the quadratic action for the
gauge fields, it does determine the mass of the 
fields in the allowed matter multiplets. In the supersymmetric case, each 
matter multiplet consists of a complex scalar field of mass 
\begin{equation}\label{eq:mass}
  M_\lambda^2 = -1+\lambda^2\ ,
\end{equation}
two Dirac fermions of mass
\begin{equation*}
m^2 = \Big(\lambda-\frac{1}{2}\Big)^2\ ,
\end{equation*}
as well as a complex scalar of mass $M_{1-\lambda}$.
These are actually short $\mathcal{N}=2$ complex supermultiplets, 
and the corresponding states in the dual CFT are $\mathcal{N}=2$ chiral primaries.\footnote{Notice that in 3 Euclidean dimensions the action of a supercharge on a real scalar field gives a Dirac fermion.}

The propagation of the free (massive) scalar or spinor fields on AdS$_3$ is 
unambiguously fixed by the respective equations of motion
provided one specifies the asymptotic behaviour of the fields at the boundary, i.e.\
the conformal dimensions of the dual superconformal fields. 
For the fields in the above mass windows, there are two natural boundary conditions
one may choose, and we shall refer to them as the $(\pm)$ quantisations;  for the
scalar fields the relevant dual conformal dimensions are again $(h,h)$ with
$\Delta=2h$ being given by  (\ref{eq:bos_scal_dim}), while for a massive Dirac fermion 
the relevant conformal dimensions are $(h+\tfrac{1}{2},h)$ and $(h,h+\tfrac{1}{2})$ with
$\Delta  = 2 h +\tfrac{1}{2}$ given by 
\begin{equation}\label{eq:qferm}
\Delta^F_+=\frac{3}{2}-\lambda\ ,\qquad
\Delta^F_-=\lambda+\frac{1}{2}\qquad \hbox{i.e.}\qquad
h_+^F=  \frac{1}{2} (1-\lambda)\ , \qquad h_-^F = \frac{\lambda}{2} \ .
\end{equation}
The contribution of the complex scalar field with $\Delta=2h$ to the 1-loop partition function is 
again  given by  (\ref{eq:scalar_pm}), while that of a 
Dirac fermion with conformal dimensions $(h+\tfrac{1}{2},h)$ 
and $(h,h+\tfrac{1}{2})$ is \cite{Creutzig:2011fe}
\begin{equation}\label{eq:spinor_pm}
 Z^\Delta_{\text{spinor}}=\prod_{m,n=0}^\infty(1+q^{h+\frac{1}{2}+m}\bar{q}^{h+n})(1+q^{h+m}\bar{q}^{h+\frac{1}{2}+n}) \ .
\end{equation}
Notice that supersymmetry determines unambiguously the quantisation of all fields in a supermultiplet in terms of, 
for instance, the quantisation of the scalar $M_\lambda^2$.
More precisely, the two scalars $M_\lambda^2$ and $M_{1-\lambda}^2$ are quantised in an opposite fashion, 
while the fermions are quantised, due to our conventions~\eqref{eq:qferm}, in the same way as $M_{1-\lambda}^2$.
This is  illustrated in 
fig.~\ref{fig:supermulti}. 
\begin{figure}[bth]
\begin{centering}
\begin{tikzpicture}[description/.style={fill=white,inner sep=2pt}]
\matrix(m)[matrix of math nodes,
row sep=1em, column sep=3em,
text height=1.5ex, text depth=0.25ex]
{
M_\lambda (+) &
m\, (-) &
M_{1-\lambda}\, (-) \\[3pt]
\makebox[25ex][c]{$\left( \frac{1}{2}+\frac{\lambda}{2},\frac{1}{2}+\frac{\lambda}{2}\right)$} &
\makebox[25ex][c]{
$2\times \begin{array}{c}
\left(\frac{\lambda}{2},\frac{1}{2}+\frac{\lambda}{2}\right)\\[3pt]
\left(\frac{1}{2}+\frac{\lambda}{2},\frac{\lambda}{2}\right)
\end{array}$
}&
{\makebox[25ex][c]{$\left(\frac{\lambda}{2},\frac{\lambda}{2}\right)$}}\\[20pt]
M_{\lambda}  (-)&
m\, (+) &
M_{1-\lambda}(+) \\[3pt]
\makebox[25ex][c]{$\left( \frac{1}{2}-\frac{\lambda}{2},\frac{1}{2}-\frac{\lambda}{2}\right)$} &
\makebox[25ex][c]{
$2\times \begin{array}{c}
\left(\frac{1}{2}-\frac{\lambda}{2},1-\frac{\lambda}{2}\right)\\[3pt]
\left(1-\frac{\lambda}{2},\frac{1}{2}-\frac{\lambda}{2}\right)
\end{array}$
}&
\makebox[25ex][c]{$\left(1-\frac{\lambda}{2},1-\frac{\lambda}{2}\right)$}\\
};
\path[<->,font=\scriptsize,>=angle 90]
(m-2-1) edge node[auto] {$Q^\pm$, $\tilde{Q}^\pm$} (m-2-2)
(m-2-2) edge node[auto] {$Q^\pm$, $\tilde{Q}^\pm$} (m-2-3)
%
%
(m-4-1) edge node[auto] {$Q^\pm$, $\tilde{Q}^\pm$} (m-4-2)
(m-4-2) edge node[auto] {$Q^\pm$, $\tilde{Q}^\pm$} (m-4-3);
\end{tikzpicture}\\
\end{centering}
\caption{
Conformal dimensions of the scalar and spinor fields 
in the two short $\mathcal{N}=2$ complex supermultiplets.
Here $Q^\pm$ and $\tilde{Q}^\pm$ are the left- and right-moving 
$\mathcal{N}=2$ supercharges in the CFT. Since the representation is short, 
one of the two supercharges of each chirality always acts trivially. The
Dirac fermions have multiplicity $2$ since the scalar fields are complex.
\label{fig:supermulti}}
\end{figure}

\smallskip

The complete matter spectrum of the higher spin theory of \cite{Creutzig:2011fe}
consists of two such ${\cal N}=2$
multiplets that are quantised again in the opposite fashion; altogether the 
1-loop partition function $\mathcal{Z}^\lambda_{\text{1-loop}}$ of this theory is therefore 

\begin{equation}
 \mathcal{Z}^\lambda_{\text{1-loop}} = 
\mathcal{Z}_{\text{gauge}} \times Z_{\text{scalars}}\times Z_{\text{spinors}} \ ,
\end{equation}
where $\mathcal{Z}_{\text{gauge}}$ was defined in (\ref{gauge1}), and the scalar and spinor 
contributions are 
\begin{align}\label{eq:pf_long}
Z_{\text{scalars}}&=
Z^{\Delta^B_+(\lambda)}_{\text{scalar}}Z^{\Delta^B_-(\lambda)}_{\text{scalar}}
Z^{\Delta^B_+(1-\lambda)}_{\text{scalar}}Z^{\Delta^B_-(1-\lambda)}_{\text{scalar}}\\ \notag
Z_{\text{spinors}}&
= \left(Z^{\Delta^F_+}_{\text{spinor}}Z^{\Delta^F_-}_{\text{spinor}}\right)^2
\ .
\end{align}
Note that the total partition function is invariant under $\lambda \mapsto 1-\lambda$. In 
terms of  ${\cal N}=2$ supermultiplets, we have 
\begin{equation}\label{eq:pf_short}
\mathcal{Z}^\lambda_{\text{1-loop}} = 
\mathcal{Z}_{\text{gauge}} \times
\mathcal{Z}^{\lambda,+}_{\text{matter}}\times  \mathcal{Z}^{\lambda,-}_{\text{matter}} \ ,
\end{equation}
where the first two factors denote the contribution of the 
two $\mathcal{N}=2$ matter multiplets from above
\begin{align}
 \mathcal{Z}^{\lambda,+}_{\text{matter}} &= 
Z^{\Delta^B_+(\lambda)}_{\text{scalar}} \,
 Z^{\Delta^B_-(1-\lambda)}_{\text{scalar}} \,
\left(Z_{\text{spinor}}^{\Delta^F_-}\right)^2
=Z^{\lambda}_{\text{scalar}}\left( Z^{\frac{1}{2}+\lambda}_{\text{spinor}}\right)^2 
Z^{1+\lambda}_{\text{scalar}}\\
 \mathcal{Z}^{\lambda,-}_{\text{matter}} &=  
Z^{\Delta^B_+(1-\lambda)}_{\text{scalar}}\,
Z^{\Delta^B_-(\lambda)}_{\text{scalar}} \,
\left(Z_{\text{spinor}}^{\Delta^F_+}\right)^2
=Z^{1-\lambda}_{\text{scalar}}\left( Z^{\frac{3}{2}-\lambda}_{\text{spinor}}\right)^2 Z^{2-\lambda}_{\text{scalar}} 
=  \mathcal{Z}^{1-\lambda,+}_{\text{matter}}  \ .
\end{align}
Expanding them out as above, we then have explicitly 
\begin{equation}\label{Zmat}
\mathcal{Z}^{\lambda,+}_{\text{matter}}  = 
\prod_{m,n=0}^\infty\frac{(1+q^{\frac{\lambda}{2}+\frac{1}{2}+m}\bar{q}^{\frac{\lambda}{2}+n})^2\,
(1+q^{\frac{\lambda}{2}+m}\bar{q}^{\frac{\lambda}{2}+\frac{1}{2}+n})^2}
{(1-q^{\frac{\lambda}{2}+m}\bar{q}^{\frac{\lambda}{2}+n})^2\,
(1-q^{\frac{\lambda}{2}+\frac{1}{2}+m}\bar{q}^{\frac{\lambda}{2}+\frac{1}{2}+n})^2} \ .
\end{equation}

\subsection{The superconformal coset}

It was  proposed in \cite{Creutzig:2011fe} that the above higher spin theory is dual to
the 't~Hooft like limit of a family of minimal ${\cal N}=2$ superconformal coset theories.
In this section we want to review the relevant superconformal field theories.

Recall that we can associate to each bosonic affine algebra $\su(N)_k$ an
${\cal N}=1$  supersymmetric affine algebra ${\mathcal \su}(N)^1_{k+N}$; 
the latter is actually isomorphic to the direct sum of the bosonic algebra $\su(N)_k$ 
together with $\dim(\su(N))$ free Majorana fermions. In analogy to this, 
we also denote by 
$\un(1)^{1}_k$ the direct sum of $\un(1)_k$ and (the chiral superalgebra of)
a single Majorana fermion. 

The cosets that are relevant for us are then 
\begin{equation}\label{eq:cosets_manifest}
\mathcal{W}_{N,k} = 
\frac{\su(N+1)^{1}_{k+N+1}}{\su(N)^{1}_{k+N+1}\oplus \un(1)^{1}_{\kappa}} \ ,
\end{equation}
where $\kappa = N(N+1)(k+N+1)$ is the `level' of the $\un(1)$ algebra 
(as defined above eq.~\eqref{eq:u1_affine_chars}).
They are manifestly $\mathcal{N}=1$ supersymmetric, but according to 
Kazama and Suzuki~\cite{Kazama:1988qp,Kazama:1988uz}, the actual chiral algebra
contains the ${\cal N}=2$ superconformal algebra. Geometrically, this is a consequence of 
the fact that the coset (\ref{eq:cosets_manifest}) is associated to the 
homogeneous space
\begin{equation}
  \mathbb{C}\mathbb{P}^N=\frac{{\rm U}(N+1)}{{\rm U}(N)\times {\rm U}(1)}\ ,
\end{equation}
which is actually a Hermitian symmetric space, i.e.\ possesses a complex structure. 
We should also mention in passing that \eqref{eq:cosets_manifest} coincides with the 
Drinfel'd-Sokolov reduction of the affine superalgebra 
$\sgl(N+1|N)_{k_{\mathrm{DS}}}$ at level \cite{Ito:1990ac}
\begin{equation}
 k_{\mathrm{DS}}=-1+\frac{1}{k+n+1}\ . 
\end{equation}

Given that the ${\cal N}=1$ superconformal algebras are actually isomorphic to direct sums
of the corresponding bosonic subalgebras and free Majorana fermions, we can reformulate
the bosonic subalgebra of $\mathcal{W}_{N,k}$ in \eqref{eq:cosets_manifest} as
\begin{equation}\label{eq:cosets}
\mathcal{W}_{N,k}^{(0)} = 
 \frac{\su(N+1)_k\oplus \so(2N)_1}{\su(N)_{k+1}\oplus \un(1)_{\kappa}}\ ,
\end{equation}
where $\so(2N)_1$ is the bosonic algebra associated to the $2N$ free Majorana fermions
that survive after subtracting from the $N^2+2N$ free fermions of the numerator 
in \eqref{eq:cosets_manifest} the $N^2$ free fermions of the denominator. The central 
charge of the coset algebra ${\cal W}_{N,k}$ is therefore 
\begin{equation}\label{cdef}
c = (N-1) + \frac{k N (N+2)}{k+N+1} - \frac{(k+1) (N^2-1)}{k+N+1} = \frac{3kN}{k+N+1} \ .
\end{equation}
In the following we shall mostly use the bosonic coset description \eqref{eq:cosets}; note that
this description contains implicitly the supersymmetry generators as long as we describe
the $\so(2N)_1$ algebra in terms of $2N$ free Majorana fermions. 
\smallskip

We shall also need to understand how the denominator of 
\eqref{eq:cosets} is embedded into the numerator. The embedding 
of $\su(N)\oplus \un(1)$ into the first factor (i.e.\ into $\su(N+1)$) is determined by the usual
embedding of ${\rm SU}(N) \times {\rm U}(1) \hookrightarrow {\rm SU}(N+1)$, 
\begin{equation}\label{i1em}
\imath_1 (v,w) = \begin{pmatrix}
   w^N  & 0\\
 0 & \bar{w}  v 
\end{pmatrix}  \in {\rm SU}(N+1) \ ,
\end{equation}
where $v \in {\rm SU}(N)$ and $w\in {\rm U}(1)$. Let us denote by $K\in \su(N+1)$ the 
image of  the $\un(1)$ Lie algebra generator
(i.e.\ $K$ is the diagonal matrix
with entries $(N,-1, \ldots, -1)$);
its OPE is then of the form
%
\begin{equation}\label{JOPE}
K(z_1) K(z_2) = \frac{k N (N+1) }{(z_1-z_2)^2} + {\cal O}(1)\ .
\end{equation}
%
In order to understand the embedding into the 
$\so(2N)$ factor, recall that we can think of $\so(2N)$ as 
the Lie algebra of the Lie group ${\rm SO}(N,N)$  of 
$2N \times 2N$ matrices $M$ satisfying $M G M^{t}=G$ with 
\begin{equation}
 G = \begin{pmatrix}
       0 & \mathbb{1_N}\\
      \mathbb{1_N} & 0
     \end{pmatrix}\ .
\end{equation}
We then embed ${\rm SU}(N) \times {\rm U}(1)\hookrightarrow {\rm SO}(N,N)$ 
(the scaling of the ${\rm U}(1)$ embedding relative to (\ref{i1em}) is fixed by ${\cal N}=1$ 
supersymmetry, see (\ref{eq:super_free_char}) below) as
\begin{equation} \label{eq:embedd_susy}
\imath_2(v,w) = 
\begin{pmatrix}
   \bar{w}^{(N+1)}\, v & 0\\
0 &  w^{(N+1)} \bar{v} 
\end{pmatrix} \in {\rm SO}(N,N) \ ,
\end{equation}
where $\bar{v}$ denotes the complex conjugate matrix to $v\in {\rm SU}(N)$. Again we denote by 
$j\in \so(2N)$ the image of the $\un(1)$ Lie algebra generator (whose first $N$ diagonal entries
are $-(N+1)$, with the remaining diagonal entries being equal to $N+1$); its OPE is
then 
\begin{equation}\label{jOPE}
j(z_1) j(z_2) = \frac{N (N+1)^2 }{(z_1-z_2)^2} + {\cal O}(1) \ .
\end{equation}
%
Together with (\ref{JOPE}) it then follows that the current
\begin{equation}
J = \frac{1}{k+N+1} \, \Bigl( K - \frac{k}{N+1}\,  j \Bigr) 
\end{equation}
is primary with respect to the denominator algebra; it therefore describes a $\un(1)$-current
of the coset algebra $\mathcal{W}_{N,k}$. It can be identified with the $\un(1)$-current of the
${\cal N}=2$ superconformal subalgebra, and with the above normalisation of $K$ and $j$ it is canonically 
normalised so that 
\begin{equation}
J(z_1) J(z_2) = \frac{c}{3 (z_1-z_2)^2} + {\cal O}(1) \ ,
\end{equation} 
where $c$ is given in (\ref{cdef}). 
\medskip

The irreducible representations of $\mathcal{W}_{N,k}$ can again 
be described in the usual manner.
Let us denote by NS the Neveu-Schwarz representation
of the $2N$ Majorana fermions. (From the point of view of $\so(2N)_1$, NS is therefore
the direct sum of the vacuum and the vector representation.) For any 
integrable representation $\Lambda\in \Yan_{N+1,k}$  of $\su(N+1)_k$ we then
consider the decomposition of the tensor product
\begin{equation}\label{eq:branching_susy_mult}
 \Lambda  \otimes \mathrm{NS} =\bigoplus_{\Xi, l} \, 
 (\Lambda;\Xi, l)\otimes
\Xi\otimes l
\end{equation}
with respect to $\su(N)_{k+1}\oplus\un(1)_\kappa$. 
Here 
$\Xi\in\Yan_{N,k+1}$ labels the representations of $\su(N)_{k+1}$, while 
$l\in\mathbb{Z}_{\kappa}$ describes the 
representations of  $\un(1)_\kappa$. In order to understand which
representations of $\su(N)_{k+1}\oplus\un(1)_\kappa$ appear in this decomposition, 
let us write $\Lambda$ and $\Xi$ in terms of the usual orthogonal basis as 
\begin{equation}
\Lambda = \sum_{j=0}^{N} \Lambda_j \varepsilon_j
- \frac{|\Lambda|}{N+1} \sum_{j=0}^{N} \varepsilon_j  \qquad \hbox{and} \qquad 
\Xi = \sum_{j=1}^{N} \Xi_j  \varepsilon_j
- \frac{|\Xi|}{N} \sum_{j=1}^{N} \varepsilon_j \ ,
\end{equation}
where $\Lambda_j$ and $\Xi_j$ are the number of boxes  in the $j$'th row 
of $\Lambda$ and  $\Xi$, respectively. (For the 
case of $\su(N+1)$ the first row is the zero'th row, while for $\su(N)$, the rows are labelled 
by $1,\ldots, N$.) Given the structure of the
embedding (\ref{i1em}), the weight of the $\un(1)_\kappa$ representation labelled by $l$ in eq.~\eqref{eq:branching_susy_mult} is then of the form
\begin{equation}
\omega_l = \frac{l}{N(N+1)} \, \Bigl( N \varepsilon_0 - \sum_{j=1}^{N} \varepsilon_j \Bigr) \qquad
\hbox{so that} \qquad 
\omega_l( K) = l \ , 
\end{equation}
while $\omega_l$ vanishes on all generators of $\su(N)_{k+1}$ under the embedding $\imath_1$. 
The root lattice of $\su(N+1)$ is generated by the vectors $\varepsilon_j$, and hence
the selection rule that $\Lambda - \Xi - \omega_l$ lies in the root lattice of $\su(N+1)$
simply means that the coefficients of all $\varepsilon_j$ are integer; for $j\neq 0$ this
is precisely the condition that 
\begin{equation}\label{eq:sel_rules_susy}
  \frac{|\Lambda|}{N+1}-\frac{|\Xi|}{N}-\frac{l}{N(N+1)}\equiv 0\mod 1\ ,
\end{equation}
and it is easy to see that then also the coefficient of $\varepsilon_0$ is integer. 
Note that (\ref{eq:sel_rules_susy}) determines $l$  in terms of 
$\Lambda$ and $\Xi$ only modulo $N(N+1)$; since $l$ is defined modulo 
$\kappa = N (N+1) (N+k+1)$, it is not completely fixed by 
\eqref{eq:sel_rules_susy}.\footnote{The level of the $\un(1)$ algebra is the central term
in the current-current OPE where the current has been normalised so that the spectrum of its
zero mode consists of  the integers. In our case the correctly normalised current is $K+j$, 
and the level can then be read off from (\ref{JOPE}) and (\ref{jOPE}).}

The multiplicity spaces labelled by 
$(\Lambda;\Xi,l)$ satisfying \eqref{eq:sel_rules_susy} then define
representations of $\mathcal{W}_{N,k}$. In fact, all representations of 
$\mathcal{W}_{N,k}$ can be described in this manner. However, not
all triplets $(\Lambda;\Xi,l)$ lead to inequivalent representations;
the relevant identification rules are worked out in \cite{Gepner:1989jq}.

\smallskip

The character of $2N$ Neveu-Schwarz Majorana fermions equals
\begin{equation}\label{eq:FF_characters_R_NS}
 \theta(q,u)= \tr_{\mathrm{NS}} q^{L_0} u= \prod_{n=0}^\infty\prod_{i=1}^{N}
(1+u_iq^{n+\frac{1}{2}})(1+\bar{u}_iq^{n+\frac{1}{2}})\ ,
\end{equation}
where  $u$ is an $\SO(N,N)$ group element with eigenvalues
$\{u_i,\bar{u}_i\}_{i=1}^{N}$. Together with
the affine characters defined in  (\ref{eq:slnk_chars}) and (\ref{eq:u1_affine_chars})
we then have the identity
\begin{equation}\label{eq:branching_susy}
  \ch^{N+1,k}_\Lambda(q,\imath_1(v,w))\, \theta(q,\imath_2(v,w))
  =\sum_{\Xi,l} 
b^{N,k}_{\Lambda;\Xi, l}(q)\, \ch^{N,k+1}_\Xi(q,v) \, \Theta^\kappa_ {l}(q,w)\ , 
\end{equation}
where 
\begin{equation}\label{eq:vir_coset_char_susy}
 b^{N,k}_{\Lambda;\Xi, l}(q)=\tr_{(\Lambda;\Xi, l)}
q^{L_0}\ ,
\end{equation}
is again the coset character. 
\smallskip

The simplest CFT is as before the charge conjugation theory whose full space of states is
of the form
\begin{equation}\label{eq:hs_KZ_min}
\mathcal{H}^{N,k}_s=\bigoplus_{[\Lambda;\Xi,l]} [\Lambda;\Xi,l] \otimes \overline{[\Lambda;\Xi,l]}\ ,
\end{equation}
where $[\Lambda;\Xi,l]$ denotes again the equivalence classes of coset representations.
The corresponding torus partition function
\begin{equation}
 \mathcal{Z}^{N,k}(q)=|q^{-\frac{c}{24}}|^2\sum_{[\Lambda;\Xi,l]}|b^{N,k}_{\Lambda;\Xi, l}(q)|^2
\end{equation}
is then modular invariant with respect to the appropriate modular group
(namely the congruence subgroup that is generated by $S$ and $T^2$).
Here we have restricted ourselves to the (unprojected) NS-NS sector.
The R-sector representations do not, in any case, contribute 
to the perturbative spectrum in the 't~Hooft limit since their conformal dimensions 
are proportional to $c$ (which goes to infinity in the limit).

\subsection{The duality}
\label{sec:duality}

As was already mentioned above,  it was 
proposed in \cite{Creutzig:2011fe} that the higher spin theory of 
section~\ref{sec:hs_sugra}  is dual to the large $N,k$ limit  (\ref{eq:t_hooft})
of the above ${\cal N}=2$ minimal model superconformal field theories. 
In order to define the limit, we restrict, as in the bosonic case of 
section~\ref{eq:non_susy_strory},  the spectrum of (\ref{eq:hs_KZ_min}) to those 
representations $(\Lambda;\Xi,l)$ for which both $\Lambda$ and $\Xi$  can be
labelled by pairs of Young diagrams ${\bf \Lambda}$ and ${\bf \Xi}$ as in 
figure~\ref{fig:inf_young}. We want to show in the following that with this restriction 
(and after removing the relevant null-vectors, see below) the partition functions between 
the two descriptions agree. This provides again very non-trivial evidence in favour of
this duality.

\subsubsection{The higher spin partition function}

Let us begin by rewriting the higher spin partition function \eqref{eq:pf_short} 
as in the bosonic case, see eq.~\eqref{eq:comparable_form}, except that now 
the relevant algebra is $\gl(\infty|\infty)_+$, rather than $\gl(\infty)_+$. In order
to do so we need to fix some conventions.

Recall that, as a vector space, the algebras $\gl(\infty)_+$ and 
$\gl(\infty|\infty)_+$ are isomorphic. The only difference is that  for the 
superalgebra $\gl(\infty|\infty)_+$  
we distinguish between the bosonic generators $E_{ij}$ for which $i+j$ is even,
and the fermionic generators $E_{ij}$ for which $i+j$ is odd. Correspondingly we then 
define commutation and anti-commutation relations. It is clear from this description that 
we have again a representation of $\gl(\infty|\infty)_+$ on $\mathbb{C}^{\mathbb{N}_0}$.

\begin{figure}[hbt]
\psfrag{i}{$i$}
\psfrag{j}{$k$}
\psfrag{k}{$j$}
\psfrag{e1}{$i\leq j,k$}
\psfrag{e22222222222222}{$i<j$ if $i$ and $j$ are odd}
\psfrag{e3}{$i<k$ if $i$ and $k$ are even}
\centering{\includegraphics{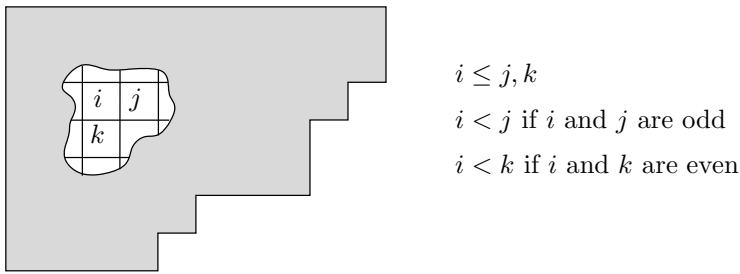}}
\caption{A supertableau of shape $\Lambda$ and type $\gl(\infty|\infty)_+$ is a 
filling of the boxes of a Young diagram $\Lambda$ with elements from $\mathbb{N}_0$ 
such that the entries of the boxes are ordered as indicated in the figure. \label{fig:supertab}}
\end{figure}

The tensor products of this fundamental representation are completely decomposable
into irreducible representations, and these are again labelled by Young diagrams~\cite{sergeev,berele}. In order
to describe the associated character of $\gl(\infty|\infty)_+$, we need to introduce 
supertableaux. A supertableau is a filling of the Young diagram 
$\Lambda$ by elements from $\mathbb{N}_0$, where the entries do not decrease along
rows and columns, and the direction in which they strictly increase depends on the cardinality
of the corresponding entries; the precise rule is explained in fig.~\ref{fig:supertab}.

We can label the basis elements of $\Lambda$ by the different Young supertableaux 
$T\in \STab_\Lambda$ of shape $\Lambda$, and the weight of $T$ is the 
sum of the fundamental weights $\epsilon_i$ associated to $T$, i.e.\ 
$\wt(T)=\sum_{i\in T} \epsilon_i$. (Note that the Cartan subalgebra 
of $\gl(\infty|\infty)_+$  can again be taken to consist of the diagonal matrices, 
and $\epsilon_i$ is then as before defined by eq.~(\ref{fundweight}).) 
The supercharacter of the  $\gl(\infty|\infty)_+$ representation labelled by $\Lambda$ is then 
\begin{equation}\label{eq:gl_inf_char_susy}
  \sch_\Lambda (e^H) = \sum_{T\in\STab_\Lambda} e^{\wt(T)(H)}\prod_{j\in T}(-1)^j\ .
\end{equation}
The generalisation of eq.~\eqref{eq:det_uinf} that is relevant for us is now 
\begin{align}\label{sdet}
\frac{1}{\sdet(1-\mathcal{U}(h)\otimes \mathcal{U}(h)^*)}=
&\prod_{m,n=0}^\infty\frac{(1+q^{h+\frac{1}{2}+m}\bar{q}^{h+n})
(1+q^{h+m}\bar{q}^{h+\frac{1}{2}+n})}
{(1-q^{h+m}\bar{q}^{h+n})(1-q^{h+\frac{1}{2}+m}\bar{q}^{h+\frac{1}{2}+n})} \notag\\
=&\, \sum_{\Lambda}\sch_\Lambda(\mathcal{U}(h))\, \, \sch_\Lambda(\mathcal{U}(h)^*)\ ,
\end{align}
where $\sdet$ denotes the superdeterminant, and $\mathcal{U}(h)$ is an $\GL(\infty|\infty)_+$ diagonal matrix with matrix elements
\begin{equation}\label{Usdef}
 \mathcal{U}(h)_{jj} = (-1)^{j} \, q^{h+\frac{j}{2}} \ .
\end{equation}
On these group elements the supercharacter reads explicitly
\begin{equation}\label{eq:ev_sch}
\sch_\Lambda(\mathcal{U}(h))=\sum_{T\in \STab_\Lambda}\prod_{i\in T}q^{h+\frac{i}{2}}\ ,
\end{equation}
since the parity signs in eq.~\eqref{Usdef} cancel against those in 
eq.~\eqref{eq:gl_inf_char_susy}.
Using exactly the same arguments as for the bosonic case, see
eq.~\eqref{eq:comb_id_1}, this allows us to write the 
partition function~\eqref{eq:pf_short}  (see in particular (\ref{Zmat}))  in the form
\begin{equation}\label{eq:susy_pf_combin_sum}
 \mathcal{Z}^\lambda_{\text{1-loop}} = 
 \mathcal{Z}_{\text{gauge}}\sum_{\Lambda_l,\Lambda_r,\Xi_l,\Xi_r}
\bigl|
\sch_{\Lambda_l}(\mathcal{U}_+) \,
\sch_{\Lambda_r}(\mathcal{U}_{+}) \,
\sch_{\Xi_l}(\mathcal{U}_{-}) \,
\sch_{\Xi_r}(\mathcal{U}_{-})
\bigr|^2\ ,
\end{equation}
where $\mathcal{U}_+=\mathcal{U}\bigl(h=\frac{\lambda}{2}\bigr)$ and $\mathcal{U}_-=\mathcal{U}\bigl(h=\frac{1-\lambda}{2}\bigr)$.
%
%

\subsubsection{The superconformal partition function}

Now we come to the CFT partition function in the 't~Hooft limit. Using the form of 
\eqref{i1em}, we can express the character of $\su(N+1)_k$ in the large $k$ limit as 
\begin{equation}\label{first}
 \ch_\Lambda^{N+1,k} \bigl(q,\imath_1(v,w)\bigr)=
\frac{q^{h^{N+1,k}_\Lambda}[\ch_\Lambda^{N+1}(\imath_1(v,w))
+\mathcal{O}(q^{k-\Lambda_1+1})]}
{\prod_{n=1}^\infty (1-q^n)^N \prod_{i\neq j=0}^N(1-v_i\bar{v}_jq^n)}  \ ,
\end{equation} 
where $\{v_i\}_{i=1}^N$ are the eigenvalues of $v\in \SU(N)$ and we have 
defined $v_0=w^{N+1}$. For the other characters in \eqref{eq:branching_susy}  
we have similarly
\begin{align}
\ch_\Xi^{N,k+1}(q,v)&=
\frac{q^{h^{N,k+1}_\Xi}[\ch_\Xi^{N}(v)+\mathcal{O}(q^{k-\Xi_1+2})]}
{\prod_{n=1}^\infty (1-q^n)^{N-1} \prod_{i\neq j=1}^N(1-v_i\bar{v}_jq^n)}\label{second} \\
\Theta^\kappa_l(q,w)& = \frac{q^{h_l^\kappa}}{\prod_{n=1}^\infty(1-q^n)}
\left[w^l+\mathcal{O}\Big(q^{\frac{\kappa}{2}-|l|}\Big)\right]\ , \qquad h_l^\kappa 
= \tfrac{l^2}{2\kappa} \ . \label{third}
\end{align}
If we define the leading term of the coset character via 
\begin{equation}\label{eq:dom_term_branch_full_susy}
  b_{\Lambda;\Xi, l}^{N,k}(q) = q^{h_\Lambda^{N+1,k}-h_\Xi^{N,k+1}- h_l^\kappa}
\left[
a_{\Lambda;\Xi, l}^{N}(q)+\mathcal{O}(q^{k-\Lambda_1+1}) 
+ {\cal O}( q^{k-\Xi_1+2})\right]\ ,
\end{equation}
it follows from eq.~\eqref{eq:branching_susy} that we have the $k$-independent identity
\begin{equation}\label{eq:branching_aux_func_susy}
\ch^{N+1}_\Lambda(\imath_1\bigl(v,w)\bigr) \, \vartheta\bigl(q,\imath_2(v,w)\bigr)
= \sum_{\Xi,l}a^{N}_{\Lambda;\Xi,l}(q)\ch^{N}_\Xi(v) \, w^l\ ,
\end{equation}
where the sum runs over all $\Xi\in\Yan_N$ and $l$ must
obey the selection rule~\eqref{eq:sel_rules_susy}. 
Note that the denominators of (\ref{first}), (\ref{second}) and (\ref{third}) cancel
among each other, except for the factors with $i=0$ or $j=0$ in (\ref{first}); 
because of (\ref{eq:embedd_susy}), these
combine with the contribution of the $N$ Dirac fermions from $\theta$
(see \eqref{eq:FF_characters_R_NS}) to the supersymmetric combination
\begin{equation}\label{eq:super_free_char}
 \vartheta(q,u)=\prod_{n=0}^\infty\prod_{i=1}^{N}
\frac{(1+u_iq^{n+\frac{1}{2}})\, (1+\bar{u}_iq^{n+\frac{1}{2}})}{
(1-u_iq^{n+1})\, (1-\bar{u}_iq^{n+1})} \ ,
\end{equation}
where $u$ is an $\SO(N,N)$ matrix with eigenvalues $\{u_i,\bar{u}_i\}_{i=1}^N$.
\smallskip

The next step of the argument consists of parametrising the different solutions for $l$
satisfying \eqref{eq:sel_rules_susy} in terms of ${\rm U}(N+1)$ and ${\rm U}(N)$ 
representations. Recall that the ${\rm U}(N)$  representations are labelled 
by pairs of Young diagrams ${\bf \Xi}=(\Xi_l,\Xi_r)$, 
where the corresponding ${\rm U}(1)$ charge is given by 
$|{\bf \Xi}|_{-} = |\Xi_r| - |\Xi_l|$. For a given ${\rm SU}(N)$
representation $\Xi$, there are different ${\rm U}(N)$ representations
${\bf \Xi}$ that restrict to $\Xi$; the ${\rm U}(1)$ charge of the various
choices for  ${\bf \Xi}$  differ by integer multiples of $N$. Since we 
may in particular take ${\bf \Xi}=(0,\Xi)$ and ${\bf \Lambda}=(0,\Lambda)$,
it follows that a solution to \eqref{eq:sel_rules_susy}  is given by taking
\begin{equation}\label{lsol}
l = N |{\bf \Lambda}|_{-} - (N+1) |{\bf \Xi}|_{-} \ .
\end{equation}
The different possible solutions for $l$ are 
then accounted for by the different choices for 
lifting $\Lambda$ and $\Xi$ to ${\rm U}(N+1)$ and ${\rm U}(N)$
representations ${\bf \Lambda}$ and ${\bf \Xi}$, respectively, and
thus \eqref{lsol} describes the most general solution. Actually, there is now
a redundancy in our description since `shifting' the separation between
$\Lambda_l$ and $\Lambda_r$ in ${\bf \Lambda}$, and between $\Xi_l$ and
$\Xi_r$ in ${\bf \Xi}$ by the same amount does not affect $l$. However,
this redundancy disappears in the large $N$ limit, as there is then a unique
way of identifying the two finite Young diagrams.

With this parametrisation in mind, we now define the $k$-independent function as 
\begin{align}
sa^{N}_{\boldsymbol{\Lambda};\boldsymbol{\Xi}}&=
a^{N}_{\boldsymbol{\Lambda}_{N+1}; \boldsymbol{\Xi}_N ,
N|\boldsymbol{\Lambda}|_- -(N+1)|\boldsymbol{\Xi}|_{-}}\ ,
 \label{eq:short_not_branch_aux}
\end{align}
so that \eqref{eq:branching_aux_func_susy} becomes
\begin{equation}\label{eq:smart_rewriting}
 \ch^{N+1}_{\boldsymbol{\Lambda}}\bigl(\imath_1(v,w)\bigr) \,
 \vartheta\bigl(q,\imath_2(v,w)\bigr)=
 \sum_{\boldsymbol{\Xi}}
 sa^N_{\boldsymbol{\Lambda};\boldsymbol{\Xi}}(q) \,
\ch^N_{\boldsymbol{\Xi}}(v\bar{w}^{N+1})
\ .
\end{equation}
Note that for $\boldsymbol{\Lambda}=0$ this identity is just
\begin{align}\label{eq:comb_dec_susy}
  \vartheta\bigl(q,\imath_2(v,w)\bigr)=&
  \prod_{n=1}^\infty\prod_{i=1}^{N}
\frac{(1+v_i\bar{w}^{N+1}q^{n-\frac{1}{2}})\,
(1+\bar{v}_iw^{N+1}q^{n-\frac{1}{2}})}{(1-v_i\bar{w}^{N+1}q^n)\,
(1-\bar{v}_iw^{N+1}q^{n})}\\
=&\sum_{\boldsymbol{\Xi}} sa^N_{0;\boldsymbol{\Xi}}(q)\,
\ch^N_{\boldsymbol{\Xi}}(v\bar{w}^{N+1}) \ .
\end{align}
In order to describe the general case from this, let us introduce the restriction coefficients
$r^{(N)}_{\boldsymbol{\Lambda}\boldsymbol{\Phi}}$
as the multiplicities with which the $\U(N)$ characters appear in the decomposition of 
$\U(N+1)$ characters
\begin{equation}\label{eq:res_coeff}
\ch^{N+1}_{\boldsymbol{\Lambda}}\bigl(\imath_1(v,w)\bigr) 
= \sum_{\boldsymbol{\Phi}}
r^{(N)}_{\boldsymbol{\Lambda}\boldsymbol{\Phi}} \, 
\ch^N_{\boldsymbol{\Phi}}(v\bar{w}^{N+1})\ ,
\end{equation}
as well as the $\U(N)$ Clebsch-Gordan coefficients
\begin{equation}
 \ch^N_{\boldsymbol{\Lambda}} \, \ch^N_{\boldsymbol{\Xi}} 
 = \sum_{\boldsymbol{\Pi}}
c^{(N)\,\boldsymbol{\Pi}}_{\boldsymbol{\Lambda}\boldsymbol{\Xi}} \, 
\ch^N_{\boldsymbol{\Pi}}  \ .
\end{equation}
Then it follows from eq.~(\ref{eq:smart_rewriting}) that we have 
\begin{equation}\label{eq:basic_branching_indentity_susy}
 sa^N_{\boldsymbol{\Lambda};\boldsymbol{\Xi}}(q) =
\sum_{\boldsymbol{\Phi},\boldsymbol{\Psi}} 
r^{(N)}_{\boldsymbol{\Lambda}\boldsymbol{\Phi}}  \,
c^{(N)\,\boldsymbol{\Xi}}_{\boldsymbol{\Phi}\boldsymbol{\Psi}} \, 
sa^N_{0;\boldsymbol{\Psi}}(q)=\sum_{\boldsymbol{\Phi},\boldsymbol{\Psi}} 
r^{(N)}_{\boldsymbol{\Lambda}\boldsymbol{\Phi}}   \,
c^{(N)\,\bar{\boldsymbol{\Psi}}}_{\boldsymbol{\Phi}\bar{\boldsymbol{\Xi}}} \,
sa^N_{0;\boldsymbol{\Psi}}(q)
\ .
\end{equation}
Generalising the combinatorial calculation of section~\ref{sec:ff_realization}, we shall 
show in section~\ref{sec:comb} that the large $N$ limit of the branching functions equals
\begin{align}\label{eq:susy_free_ch}
sa_{0;0}(q)  =  
\lim_{N\to\infty}sa^N_{0;0}(q)&= \prod_{n=1}^\infty\frac{(1+q^{\frac{n}{2}+1})^{2n}}{(1-q^n)^{2n-1}}\\ 
sa_{0;\boldsymbol{\Xi}}(q) =  \lim_{N\to\infty}sa^N_{0;\boldsymbol{\Xi}}(q)
&= sa_{0;0}(q)\, \sch_{\Xi^t_l}(\mathcal{U}_1)\, \sch_{\Xi^t_r}(\mathcal{U}_1)\ ,
\label{eq:susy_f_tens}
\end{align}
where the $\GL(\infty|\infty)_+$ supercharacters $\sch_{\Xi}$ were defined in \eqref{eq:ev_sch}
and $\mathcal{U}_1=\mathcal{U}\big(h=\frac{1}{2}\big)$. Notice that $|sa_{0;0}(q)|^2=\mathcal{Z}_{\text{gauge}}(q)$.
\smallskip

In the final step of the argument we have to remove the null states that appear in the
limit. By analogy with the bosonic case, we propose that this amounts to replacing 
the restriction and Clebsch-Gordan coefficients by 
\begin{align}
\lim_{N\to\infty}c^{(N)\,\bar{\boldsymbol{\Psi}}}_{\boldsymbol{\Phi}\bar{\boldsymbol{\Xi}}} 
& \rightarrow 
c^{\ \ \; \Psi_r}_{\Phi_l\Xi_r}\, c^{\ \ \ \Psi_l}_{\Phi_r\Xi_l} \label{eq:cg_null} \\
\label{eq:restr_null}
\lim_{N\to\infty}r^{(N)}_{\boldsymbol{\Lambda}\boldsymbol{\Phi}} &
\rightarrow  r_{\Lambda_l\Phi_l}\, r_{\Lambda_r\Phi_r} \ .
\end{align}
Here $c_{\Phi\Psi}^{\phantom{\Phi}\Xi}$  are the Clebsch-Gordan coefficients 
of $\gl(\infty)_+$ that already appeared in section~\ref{eq:non_susy_strory}; as is explained
in appendix~\ref{app:gl}, we can also interpret them as $\gl(\infty|\infty)_+$ Clebsch-Gordan 
coefficients, i.e.\ they satisfy
\begin{equation}\label{eq:lr_susy}
\sch_\Lambda \sch_\Xi = \sum_{\Pi\in \Yan} c_{\Lambda\Xi}^{\phantom{\Lambda}\Pi}\,
\sch_\Pi\ .
\end{equation}
The coefficients appearing on the right hand side of eq.~\eqref{eq:restr_null} are the
restriction coefficients for $\gl(\infty)_+$ that can be expressed in terms of the
Clebsch-Gordan coefficients as \cite{king}
\begin{equation}\label{eq:expl_res}
r_{\Lambda\Xi}=c^{\phantom{\Xi|\Lambda/}\Lambda}_{\Xi |\Lambda/\Xi|}\ ,
\end{equation}
where $|\Lambda/\Xi|$ denotes the Young diagram with a single row of
$|\Lambda|-|\Xi|$ boxes.
In particular, eq.~\eqref{eq:expl_res} implies that $r_{\Lambda\Xi}$ can only be either 
$0$ or $1$. The coefficients  $r_{\Lambda\Xi}$ also define restriction 
coefficients for $\gl(\infty|\infty)_+$, since we have the identity
(see appendix~\ref{app:gl} for a detailed derivation)
\begin{equation}\label{eq:res_sch}
\sch_\Lambda(\mathcal{U}_0)=\sum_{\Xi\in \Yan} r_{\Lambda\Xi} \, \sch_{\Xi^t} (\mathcal{U}_1)\ ,
\end{equation}
were $\mathcal{U}_0=\mathcal{U}(h=0)$. Note that we are considering here the branching
rules of $\gl(\infty|\infty)_+$-rep\-re\-sen\-ta\-tions into representations of
the subalgebra of infinite matrices whose first row and column is zero; 
the latter algebra is again $\gl(\infty|\infty)_+$, but with a shifted definition of parity,
and this is the origin of  the transposition of $\Xi$ on the right-hand-side.
\smallskip

With these preparations we can now finally compute the partition function for the 
Hilbert space
\begin{equation}\label{eq:hs_susy}
\mathcal{H}^\lambda_s = \bigoplus_{\Lambda,\Xi\in\mathbf{Y}}
(\boldsymbol{\Lambda};\boldsymbol{\Xi})_s\otimes 
\overline{(\boldsymbol{\Lambda};\boldsymbol{\Xi})}_s
\end{equation}
of the Kazama-Suzuki coset~\eqref{eq:cosets} in the 't~Hooft limit 
\eqref{eq:t_hooft}, where we have denoted by
$(\boldsymbol{\Lambda};\boldsymbol{\Xi})_s$ the large $N$, $k$ limit of  
the $\mathcal{W}_{N,k}$ representations 
\begin{equation}
(\boldsymbol{\Lambda};\boldsymbol{\Xi})_s 
= \lim_{N,k\to\infty}\Bigl(\boldsymbol{\Lambda}_{N+1}\,;\,\boldsymbol{\Xi}_N,\,
\bigl[N|\boldsymbol{\Lambda}|_--(N+1)|\boldsymbol{\Xi}|_- \bigr]\Bigr)
\end{equation}
using the same notation as in \eqref{eq:short_not_branch_aux}. Their characters
can be computed from \eqref{eq:dom_term_branch_full_susy}, and dropping 
the null-states as in (\ref{eq:restr_null}) and (\ref{eq:cg_null}) we obtain
\begin{align}
\mathrm{Tr}_{(\boldsymbol{\Lambda};\boldsymbol{\Xi})_s}q^{L_0}&=
q^{\frac{\lambda}{2}(|\boldsymbol{\Lambda}|-|\boldsymbol{\Xi}|)}
\sum_{\boldsymbol{\Phi}, \boldsymbol{\Psi}}
r_{\Lambda_l\Phi_l}r_{\Lambda_r\Phi_r}
c^{\ \ \; \Psi_r}_{\Phi_l\Xi_r}c^{\ \ \ \Psi_l}_{\Phi_r\Xi_l}
sa_{0;0}(q)\sch_{\Psi^t_l}(\mathcal{U}_1)\sch_{\Psi^t_r}(\mathcal{U}_1) \notag
\\
{}&=
q^{\frac{\lambda}{2}(|\boldsymbol{\Lambda}|-|\boldsymbol{\Xi}|)} sa_{0;0}(q)\sch_{\Lambda_l}(\mathcal{U}_0)\sch_{\Lambda_r}(\mathcal{U}_0)
\sch_{\Xi^t_l}(\mathcal{U}_1) \sch_{\Xi^t_r}(\mathcal{U}_1) \notag\\
{}&= sa_{0;0}(q)\, \sch_{\Lambda_l}(\mathcal{U}_+)\,
\sch_{\Lambda_r}(\mathcal{U}_+)\,
\sch_{\Xi^t_l}(\mathcal{U}_-)\,
\sch_{\Xi^t_r}(\mathcal{U}_-) \ ,
\end{align}
where the $\GL(\infty|\infty)_+$ matrices $\mathcal{U}_\pm$ have been defined 
in \eqref{Usdef}, and we have used that 
\begin{equation}
\lim_{N,k\rightarrow \infty} 
\Bigl[ h_\Lambda^{N+1,k} - h_\Xi^{N,k+1} 
- \frac{(N|\boldsymbol{\Lambda}|_--(N+1)|\boldsymbol{\Xi}|_-)^2}{2 N (N+1) (N+k+1)}  \Bigr]
= \frac{\lambda}{2} \bigl( |{\bf \Lambda}| - |{\bf \Xi}| \bigr) \ . 
\end{equation}
Finally, summing over the different representations, we get
\begin{equation}
\mathrm{Tr}_{\mathcal{H}_s^\lambda}q^{L_0}\bar{q}^{\bar{L}_0}= \mathcal{Z}_{\text{gauge}}
\sum_{\boldsymbol{\Lambda},\boldsymbol{\Xi}}
\bigl|
\sch_{\Lambda_l}(\mathcal{U}_+)
\sch_{\Lambda_r}(\mathcal{U}_{+})
\sch_{\Xi^t_l}(\mathcal{U}_{-})
\sch_{\Xi^t_r}(\mathcal{U}_{-})
\bigr|^2\ ,
\end{equation}
which reproduces indeed the partition function~\eqref{eq:susy_pf_combin_sum}
of the higher spin theory.

\subsection{Free field realisation}
\label{sec:comb}

Thus we are left with proving the combinatorial identities 
\eqref{eq:susy_free_ch} and \eqref{eq:susy_f_tens}; this can be done
as in the bosonic case using free fields.

First we note that the right-hand-side of \eqref{eq:comb_dec_susy} equals
the partition function of $N$ complex fermions and $N$ complex bosons,
transforming in the fundamental and anti-fundamental representations of $\U(N)$.
More specifically, let us denote by $\psi^1,\ldots,\psi^{N}$ and $\jmath^1,\ldots,\jmath^N$
the fermionic and bosonic modes in the fundamental
representation of ${\rm U}(N)$, respectively; their complex conjugates, 
$\bar{\psi}^1,\ldots,\bar{\psi}^N$ and $\bar{\jmath}^1,\ldots,\bar{\jmath}^N$
then transform in the anti-fundamental representation. 
The full Fock space is spanned by the states of the form 
\begin{equation}\label{eq:fock_susy}
\prod_{j=1}^{n_{\bar{\psi}}}\bar{\psi}^{a_j}_{-r_j-\frac{1}{2}}\prod_{k=1}^{n_\psi} \psi^{b_k}_{-s_k-\frac{1}{2}}
\prod_{l=1}^{n_{\bar{\jmath}}} \bar{\jmath}^{c_l}_{-t_l-1}\prod_{m=1}^{n_{\jmath}} \jmath^{d_m}_{-u_m-1}\Omega \ ,
\end{equation}
where the mode numbers $r_j,s_k,t_l,u_m$ are non-negative integers. In order
to determine \eqref{eq:susy_free_ch} and \eqref{eq:susy_f_tens} we need to
count the multiplicities with which a specific $\U(N)$ representation appears in the
Fock space. 

We begin again by counting the states that transform in the trivial representation. 
By the fundamental theorem of classical invariant theory~\cite{weyl}, these states are 
linear combinations of the `basic' invariants
\begin{align}\label{eq:inv_field_susy}
& \prod_{r,s=0}^\infty 
\left(\sum_{a=1}^N \bar{\psi}^a_{-r-\frac{1}{2}}\psi^a_{-s-\frac{1}{2}}\right)^{K_{rs}} \,
\prod_{t,u=0}^\infty \left(\sum_{a=1}^N \bar{\jmath}^a_{-t-1}\jmath^a_{-u-1}\right)^{L_{tu}}\notag \\
& \qquad \times \prod_{t,s=0}^\infty 
\left(\sum_{a=1}^N \bar{\jmath}^a_{-t-1}\psi^a_{-s-\frac{1}{2}}\right)^{P_{ts}} \,
\prod_{r,u=0}^\infty \left(\sum_{a=1}^N \bar{\psi}^a_{-r-\frac{1}{2}}\jmath^a_{-u-1}\right)^{Q_{ru}} 
\, \Omega \ , 
\end{align}
where only finitely many multiplicities $K_{rs}$, $L_{tu}$, $P_{ts}$, $Q_{ru}$ are non-zero.
Note that each $P_{ts}$ and $Q_{ru}$ can only be $0$ or $1$ because the fermionic 
invariants $\sum_a \bar{\jmath}^a\psi^a$ and $\sum_a \bar{\psi}^a\jmath^a$ square to zero. 
In the $N\to\infty$ limit all the states in \eqref{eq:inv_field_susy}
are linearly independent, and it is straightforward to count them,
leading to eq.~\eqref{eq:susy_free_ch}
\begin{align}\notag
sa_{0;0}(q)&=\prod_{r,s=0}^\infty \sum_{K=0}^\infty q^{(r+s+1)K}
\prod_{t,u=0}^\infty \sum_{L=0}^\infty q^{(t+u+2)L}
\prod_{t,s=0}^\infty \sum_{P=0}^1 q^{(t+s+\frac{3}{2})P} 
\, \prod_{r,u=0}^\infty \sum_{Q=0}^1 q^{(r+u+\frac{3}{2})Q} \notag \\
& =\prod_{n=1}^\infty\frac{1}{1-q^n}\prod_{i,j=0}^\infty\frac{(1+q^{i+j+\frac{3}{2}})^2}{(1-q^{i+j+2})^2}\\
{}&=\prod_{n=1}^\infty\frac{1}{1-q^n}\prod_{s=2}^\infty \prod_{n=s}^\infty \frac{(1+q^{n-\frac{1}{2}})^2}{(1-q^n)^2}=
\prod_{n=1}^\infty\frac{(1+q^{n+\frac{1}{2}})^{2n}}{(1-q^n)^{2n-1}}\ .
\label{eq:proof_inv_s}
\end{align}
\smallskip

Finally, we need to count the multiplicity with which a specific ${\bf \Xi}$ representation of 
${\rm U}(N)$ appears; again, the argument follows the same logic as in the bosonic
calculation in section~\ref{sec:ff_realization}.
Let us consider the subspace of states of the form \eqref{eq:fock_susy} with a fixed number
of modes 
$n_\psi$, $n_{\bar{\psi}}$, $n_{\jmath}$, $n_{\bar{\jmath}}$.
Then a $\U(N)$ tensor of shape 
$\boldsymbol{\Xi}=(\Xi_l,\Xi_r)$ such that $|\Xi_l|=n_{\bar{\psi}}+n_{\bar{\jmath}}$ 
and $|\Xi_r|=n_\psi+n_\jmath$ will appear with multiplicity $0$ or $1$.
The multiplicity will be precisely $1$ if   there is (i) a Young supertableau of shape 
$\Xi_l$ with bosonic or even entries from $\{2t_l+2\}$, and fermionic or odd entries from 
$\{2 r_j+1\}$; and (ii) a Young supertableau of shape $\Xi_r$ with bosonic 
or even entries from $\{2u_m+2\}$, and fermionic or odd entries from 
$\{2 s_k+1\}$.\footnote{ The form of these entries is twice the conformal dimension of the 
modes in \eqref{eq:fock_susy}.}
This is equivalent to the requirement that there are Young supertableaux of shape 
$\Xi^t_l$ and $\Xi_r^t$ with entries from $\{2r_j\}\cup\{2t_l+1\}$ and 
$\{2s_k\}\cup\{2u_m+1\}$, respectively. Summing over all possible mode numbers and different
$n_\psi$, $n_{\bar{\psi}}$, $n_{\jmath}$, $n_{\bar{\jmath}}$ such that 
$n_\psi+n_\jmath=|\Xi_r|$  and $n_{\bar{\psi}}+n_{\bar{\jmath}}=|\Xi_l|$, their contribution
to the branching function~\eqref{eq:susy_f_tens} can be written with the help of \eqref{eq:ev_sch} in the compact form
\begin{equation}\label{eq:first_tens_susy}
\sch_{\Xi^t_l}(\mathcal{U}_1)\sch_{\Xi^t_r}(\mathcal{U}_1)\ . 
\end{equation}
Multiplying these minimal states with all the invariant states~\eqref{eq:inv_field_susy}, 
one generates all states transforming in  $\boldsymbol{\Xi}$ in the Fock space.
Thus the branching function \eqref{eq:susy_f_tens} is indeed just the product of 
\eqref{eq:first_tens_susy} with \eqref{eq:proof_inv_s}.

\section{Conclusions}

In this paper we have given strong evidence in favour of the supersymmetric
higher spin duality that was proposed in \cite{Creutzig:2011fe}. In particular,
we have shown that the 1-loop partition function of the supersymmetric higher
spin theory on AdS$_3$ can be reproduced from the 't~Hooft limit of the
dual ${\cal N}=2$ Kazama-Suzuki models. Our analysis follows in spirit
closely \cite{Gaberdiel:2011zw}, where the corresponding consistency check
for the original bosonic duality of \cite{Gaberdiel:2010pz} was performed. The main
technical advance is that we have managed to determine the branching functions
(both in the bosonic as well as the supersymmetric case) from first principles,
using a free field description. This point of view also sheds light on the origin of the 
underlying $\gl(\infty)$ symmetry (resp.\ $\gl(\infty|\infty)$ for the supersymmetric 
case)  of the partition function.

In order to make sense of the limit theory (and to match with the AdS gravity answer)
we have assumed by analogy with the bosonic case that certain CFT states become 
null and decouple in the 't~Hooft limit. It would be interesting to check
this directly (at least for the first few cases) 
by performing a similar analysis to what was done in \cite{Gaberdiel:2011zw}. In
order to be able to perform this analysis, it will be important to understand the underlying 
symmetry algebra --- i.e.\ the supersymmetric analogue of $W_\infty[\lambda]$, see 
\cite{Gaberdiel:2011wb} --- in more detail. This would also allow for a more detailed
test of the correspondence by comparing eigenvalues of the various higher spin
zero modes. We hope to come back to these issues elsewhere.

\subsection*{Acknowledgements} The work of CC and MRG is supported in parts by the
Swiss National Science Foundation. We thank Maximilian Kelm and Carl Vollenweider
for useful discussions.

\appendix

\section{Identities for $\gl(\infty|\infty)_+$}
\label{app:gl}

In this appendix we want to prove (\ref{eq:lr_susy}) as well as 
(\ref{eq:expl_res}) and \eqref{eq:res_sch}.

Let $\Sym_n$ be the group of permutations of the integers $\{1,2,\dots,n\}$.
Its irreducible representations $S_\Lambda$ are indexed by partitions of $n$, that is Young diagrams $\Lambda$ with $|\Lambda|=n$.
When restricted to  the subgroup $\Sym_m\times \Sym_n\subset \Sym_{m+n}$, a representation $S_\Lambda$ of $\Sym_{m+n}$ decomposes as~\cite[ch.~1]{macdonald}
\begin{equation}\label{eq:def_lr}
\res S_\Lambda \cong \bigoplus_{\Xi,\Pi} c_{\Xi\Pi}^{\phantom{\Xi}\Lambda}\ S_\Xi \boxtimes S_\Pi\ ,
\end{equation}
where the sum is over partitions $\Xi$ of $m$ and partitions $\Pi$ of $n$, and we used the 
symbol $\boxtimes$ to denote the tensor product between representations of different groups.

Let  $V$ be the fundamental representation of $\U(M|N)$, and $V_\Lambda$ the 
irreducible $\U(M|N)$ covariant tensor of shape $\Lambda$. 
Clearly, the two groups $\Sym_n$ and $\U(M|N)$ act naturally on the tensor product 
$V^{\otimes n}$. Because  the action of $\Sym_n$ commutes with the action of  
$\U(M|N)$, one can consider $V^{\otimes n}$ as a representation of the
product group $\Sym_n\times \U(M|N)$.
With respect to this latter action, one has the following decomposition into irreducible representations~\cite{sergeev}
\begin{equation}\label{eq:schur_weyl}
V^{\otimes n} \cong \bigoplus_\Lambda S_\Lambda \boxtimes V_\Lambda\ ,
\end{equation}
where the sum runs over all partitions $\Lambda$ of $n$ that fit into a hook with arm 
width $M$ and leg width $N$~\cite{berele}.
We shall call these partitions hook-shaped.
This type of multiplicity free decomposition is known in the mathematical literature as 
a Schur-Weyl duality.

Consider now the decomposition of the representation $V^{\otimes m}\otimes V^{\otimes n}$  
with respect to the product group $\Sym_m\times \Sym_n\times \U(M|N)$.
Applying eq.~\eqref{eq:schur_weyl}, on the one hand, to the whole tensor product 
$V^{\otimes (m+n)}$ and, on the other, to each factor $V^{\otimes m}$ and $V^{\otimes n}$
separately, one arrives at
\begin{equation}
\bigoplus_{\Lambda}\res S_\Lambda\boxtimes V_\Lambda = \bigoplus_{\Xi,\Pi} S_\Xi\boxtimes S_\Pi\boxtimes \left(V_\Xi\otimes V_\Pi\right)\ .
\end{equation}
Decomposing the restricted representation into irreducibles as in eq.~\eqref{eq:def_lr}, 
we conclude that the tensor product of irreducible $\U(M|N)$ representations must be
\begin{equation}
V_\Xi\otimes V_\Pi \cong \bigoplus_\Lambda c_{\Xi\Pi}^{\phantom{\Xi}\Lambda}\ V_\Lambda\ ,
\end{equation}
where all partitions are hook-shaped.
Setting $M=N$ and taking $N\to\infty$ we  arrive at eq.~\eqref{eq:lr_susy}.
Note that the restriction on the hook-shape disappears in this limit.
\smallskip

Finally, we want to prove (\ref{eq:expl_res}) and (\ref{eq:res_sch}), following MacDonalds's 
book \cite{macdonald} on symmetric functions.
Let $X,Y\in\GL(\infty|\infty)_+$ be two diagonal matrices, whose entries we label as 
\begin{equation}
\begin{array}{cc}
X_{2i,2i}=x_{i+1}\ ,& X_{2i+1,2i+1}=\xi_{i+1}\\
Y_{2i,2i}=y_{i+1}\ ,& Y_{2i+1,2i+1}=\eta_{i+1}
\end{array}
\ , \qquad i \in\mathbb{N}_0 \ .
\end{equation}
We define a Schur type symmetric function by
\begin{equation}\label{eq:sf_sch}
s_\Lambda(x|\xi)=\sch_\Lambda(X)=\sum_{T\in\STab_\Lambda}\prod_{j\in T}X_{jj}(-1)^j\ ,
\end{equation}
where $x=(x_1,x_2,\dots)$, $\xi=(\xi_1,\xi_2,\dots)$ are treated as formal indeterminate 
variables. Note that if we restricted  the values of $x$ and $\xi$ by setting $x_i=0$ for 
$i>M$ and $\xi_j=0$ for $j>N$,  then~\eqref{eq:sf_sch} becomes a $\U(M|N)$ character.
From~\eqref{eq:gl_inf_char_susy} and the definition of Young supertableaux
in fig.~\ref{fig:supertab} it follows that
\begin{equation}\label{eq:sf_trans}
s_\Lambda(x|\xi)=s_{\Lambda^t}(-\xi|-x)\ .
\end{equation}
Denoting $y=(y_1,y_2,\dots)$ and $\eta=(\eta_1,\eta_2,\dots)$, we can now rewrite 
\eqref{sdet} as
\begin{equation}
\prod_{i,j}\frac{(1-x_i\eta_j)(1-y_i\xi_j)}{(1-x_iy_j)(1-\xi_i\eta_j)}
=\sum_\Lambda s_\Lambda(x|\xi) s_\Lambda(y|\eta)\ ,
\end{equation}
where the left hand side is to be understood as a generating function. Next we repeat 
the argument of \cite[p. 40--41]{macdonald}.
Let us introduce a third set of independent variables  $z=(z_1,z_2,\dots)$, 
$\zeta=(\zeta_1,\zeta_2,\dots)$, and consider the product which we can rewrite
in two different ways as 
\begin{align}
& \prod_{i,j}\frac{(1-z_i\xi_j)(1-z_i\eta_j)(1-\zeta_ix_j)(1-\zeta_iy_j)}{(1-z_ix_j)
(1-z_iy_j)(1-\zeta_i\xi_j)(1-\zeta_i\eta_j)} = 
\sum_{\Lambda} s_\Lambda(z|\zeta)\, s_\Lambda(x\cup y|\xi\cup\eta) \notag \\
& \qquad =\sum_{\Xi,\Pi}
s_\Xi(z|\zeta)\, s_\Xi(x|\xi)\, s_{\Pi}(z|\zeta)\, s_\Pi (y|\eta)\\
&\qquad = \sum_{\Lambda,\Xi,\Pi}s_{\Lambda}(z|\zeta)
\Big(c_{\Xi\Pi}^{\phantom{\Xi}\Lambda}\, s_\Xi(x|\xi)\, s_\Pi(y|\eta)\Big) \ , \notag
\end{align}
where we have used (\ref{eq:lr_susy}) in the last line; this leads to the 
important relation
\begin{equation}\label{eq:br_gen}
s_\Lambda(x\cup y|\xi\cup\eta) = \sum_{\Xi,\Pi}c_{\Xi\Pi}^{\phantom{\Xi}\Lambda}\,
s_\Xi(x|\xi)\, s_\Pi(y|\eta)\ .
\end{equation}
We now specialise to $y=(w,0,0,\dots)$ and $\eta=(0,0,\dots)$. 
Then $s_\Pi(y|\eta)$ becomes a $\U(1)$ character, which is only non-zero provided that 
$\Pi$ has a single row, in which case it equals $w^{|\Pi|}$. 
Next, we choose $x$ and $\xi$ so that for all  $i\in\mathbb{N}_0$
\begin{align}\label{eq:u1_spec}
\xi_{i+1}&=(\mathcal{U}_1)_{2i,2i}=q^{i+\frac{1}{2}}\ ,& x_{i+1}&
=(\mathcal{U}_1)_{2i+1,2i+1}=-q^{i+1} \ ,&
\end{align}
where $\mathcal{U}_1$ is, as before, defined 
by $\mathcal{U}_1= \mathcal{U}(h=\tfrac{1}{2})$ and we recall that $\mathcal{U}(h)_{jj}=(-1)^j q^{h+\frac{j}{2}}$, see
eq.~\eqref{Usdef}. Then the eigenvalues of $\mathcal{U}_0=\mathcal{U}(h=0)$ are
\begin{align}\label{eq:u0_spec}
(\mathcal{U}_0)_{00}&=1\ ,& (\mathcal{U}_0)_{2i+2,2i+2}&=-x_{i+1}=q^{i+1}\ ,&  
(\mathcal{U}_0)_{2i+1,2i+1}&=-\xi_{i+1}=-q^{i+\frac{1}{2}} \ ,&
\end{align}
where again $i\in\mathbb{N}_0$.
Setting $w=1$, it follows from (\ref{eq:u0_spec}), (\ref{eq:sf_trans}), 
(\ref{eq:br_gen})  and (\ref{eq:u1_spec}) that
\begin{equation}
\sch_\Lambda(\mathcal{U}_0)=s_\Lambda (-x\cup \{1\} |-\xi)=
\sum_{\Xi}c_{|\Lambda/\Xi|\Xi}^{\phantom{|\Lambda/\Xi|}\Lambda} \, \, s_{\Xi^t}(\xi|x)=
\sum_{\Xi}c_{|\Lambda/\Xi|\Xi}^{\phantom{|\Lambda/\Xi|}\Lambda} \, \, \sch_{\Xi^t}(\mathcal{U}_1) \ .
\end{equation}
This completes the proof of (\ref{eq:expl_res}) and (\ref{eq:res_sch}).



\end{document}